\documentclass[twocolumn]{aastex62}
\usepackage{comment}

\usepackage{amsmath}
\usepackage{mathrsfs} 
\usepackage{graphicx}
\usepackage{ulem}
\usepackage{macros_vdb}

\received{15 February, 2019}
\revised{12 April 2019}
\accepted{22 April 2019}

\shorttitle{On the Orbital Decay of Globular Clusters in NGC1052-DF2}
\shortauthors{Dutta Chowdhury et al.}

\begin{document}

\title{On the Orbital Decay of Globular Clusters in NGC1052-DF2: Testing a Baryon-Only Mass Model}

\correspondingauthor{Dhruba Dutta Chowdhury}
\email{dhruba.duttachowdhury@yale.edu}

\author[0000-0003-0250-3827]{Dhruba Dutta Chowdhury}
\affil{Department of Astronomy, Yale University \\
52 Hillhouse Avenue, New Haven, CT-06511, USA}

\author[0000-0003-3236-2068]{Frank C. van den Bosch}
\affil{Department of Astronomy, Yale University \\
52 Hillhouse Avenue, New Haven, CT-06511, USA}

\author{Pieter van Dokkum}
\affil{Department of Astronomy, Yale University \\
52 Hillhouse Avenue, New Haven, CT-06511, USA}

\begin{abstract}
The dark matter content of the ultra diffuse galaxy NGC1052-DF2, as inferred from globular cluster (GC) and stellar kinematics, carries a considerable amount of uncertainty, with current constraints also allowing for the complete absence of dark matter. We test the viability of such a scenario by examining whether in a `baryon-only' mass model, the observed GC population experiences rapid orbital decay due to dynamical friction. Using a suite of 50 multi-GC $N$-body simulations that match observational constraints on both the stellar component of NGC1052-DF2 and its GC population but differ in the initial line-of-sight positions and the tangential velocities of the GCs, we show that there is a substantial amount of realization-to-realization variance in the evolution of the GCs. Nevertheless, over $\sim 10\ \rm Gyr$, some of the GCs experience significant orbital evolution. Others evolve less. A combination of reduced dynamical friction in the galaxy core and GC-GC scattering keeps the GCs afloat, preventing them from sinking all the way to the galaxy center. While the current phase-space coordinates of the GCs are not unlikely for a baryon-only mass model, the GC system does evolve over time. Therefore, if NGC1052-DF2 has no dark matter, some of its GCs must have formed further out, and the GC system must have been somewhat more extended in the past. The presence of a low mass cuspy halo, while allowed by the kinematics, seems improbable as significantly shorter inspiral timescales in the central region would quickly lead to the formation of a nuclear star cluster.
\end{abstract}

\keywords{galaxies: individual (NGC1052-DF2) -- galaxies: kinematics and dynamics}

\section{Introduction} 
\label{sec:intro}

Ultra diffuse galaxies (UDGs) are a recently discovered population of faint galaxies with large sizes. First observed in the Coma cluster by \citet{vandokkum15}, UDGs are defined as galaxies having central surface brightness, $\mu (g,0)>24\ \rm mag/arcsec^2$ and effective radius, $R_\rme > 1.5\kpc$. Since then, they have been found in diverse environments varying from low mass groups to rich galaxy clusters \citep[e.g.,][]{mihos15, koda15, delgado16, vanderburg16, janssens17, lee17, roman17a, roman17b, trujillo17, vanderburg17, cohen18, zaritsky19}. 

How UDGs fit within the overall framework of galaxy formation is not well understood. One possibility is that they are dwarf galaxies residing in halos with higher than average angular momentum and/or lower than average concentration, which endows them with larger than average sizes \citep{yozin15, amorisco16, rong17}. Other postulated formation channels include expansion of classical dwarfs ($R_\rme < 1.5\kpc$) due to supernova feedback-driven gas outflows \citep{dicinto17, chan18} or tidal heating and stripping \citep{carleton19, jiang18}. 

Dynamical masses of some UDGs have been estimated from the kinematics of their globular cluster (GC) populations \citep[e.g.,][]{beasley16b, toloba18}. Under the assumption of a Navarro-Frenk-White (NFW) halo \citep{navarro97}, these studies have yielded halo masses varying from $10^{11}-10^{12}\Msun$ and dark-to-stellar mass fractions varying from $100-1000$. The total mass in GCs ($M_{\rm GCS}$) has also been used to infer the halo mass ($M_{\rm halo}$) assuming a linear relationship between the two \citep[e.g.,][]{harris17}. While UDGs on average have larger GC populations than classical dwarfs at the same luminosity \citep[][but see also \citealt{prole19}]{beasley16a, peng16, vandokkum16, vandokkum17, lim18}, there is also a considerable amount of scatter in their GC richness \citep{lim18}. Consequently, halo masses inferred from the $M_{\rm halo}-M_{\rm GCS}$ relation (under the assumption of zero scatter) exhibit a significant amount of scatter: UDGs could be over-massive, under-massive or have the same halo mass as that expected from the extrapolation of standard subhalo abundance matching relations \citep[e.g.,][]{moster10, behroozi13} to lower stellar masses. However, it is also plausible that the relation between $M_{\rm halo}$ and $M_{\rm GCS}$ has substantial scatter at low halo masses \citep[e.g.,][]{forbes18} such that $M_{\rm GCS}$ cannot be used as a reliable halo mass indicator \citep[but see also][]{burkert19}.

\begin{figure*}
    \centering
    \includegraphics[width=\textwidth]{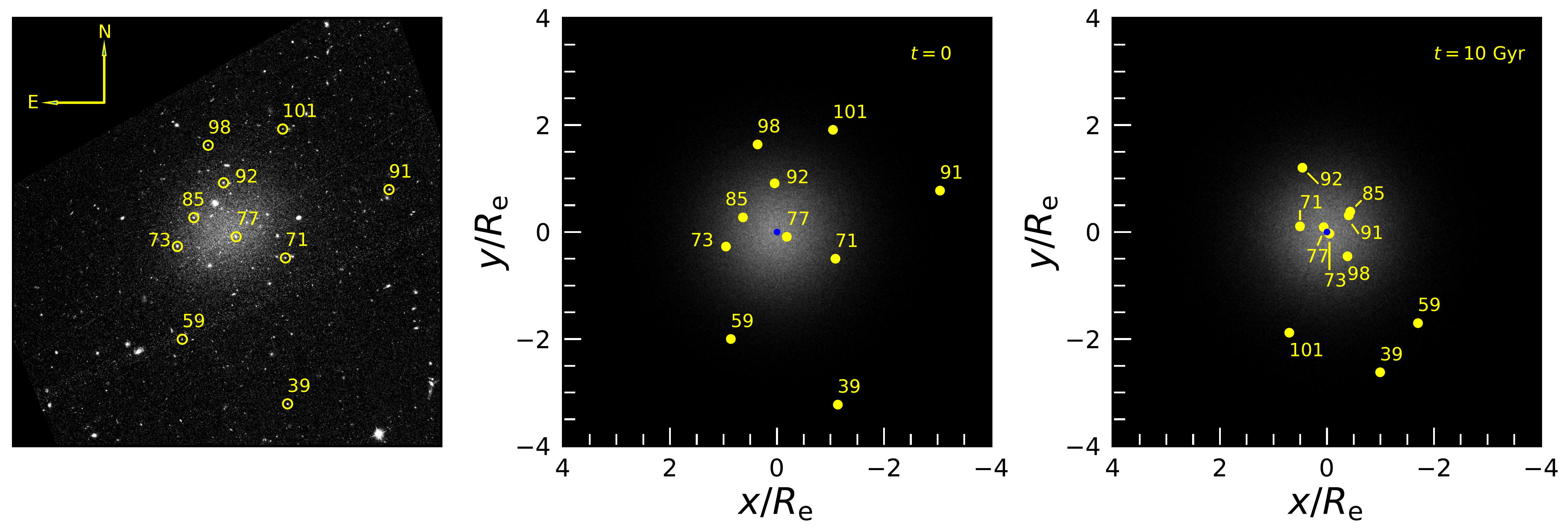}
    \label{fig:image}
    \caption{Left-hand panel: Hubble Space Telescope $V_{606}$-band image of NGC1052-DF2. The image spans $3 \arcmin \times 3 \arcmin$ or $17.6\kpc \times 17.6\kpc$ at the assumed distance of 20 Mpc. The locations of the 10 GCs considered in this paper are marked with yellow circles. Middle-panel: $x$-$y$ projection of the baryon-only $N$-body model for NGC1052-DF2 at $t=0$, which is constructed to match the observed surface brightness profile. Yellow dots indicate the GC particles, which have the same projected positions (and los velocities) as in the data, while the blue dot marks the galaxy center. Right-hand panel: Same as the middle panel but after 10 Gyr of evolution, taken from one of our 50 multi-GC simulations. Note how compared to its configuration at $t=0$, the GC system has become more compact as a consequence of orbital decay due to dynamical friction.}
\end{figure*}

In this context, the UDG NGC1052-DF2 (DF2 hereafter), located in the NGC 1052 group, is an interesting find. DF2 has a rich system of compact objects (see left-hand panel of Figure~\ref{fig:image}) with properties similar to that of $\omega$ Centauri, the brightest and the largest GC in the Milky Way \citep{vandokkum18b}. Applying the \cite{harris17} $M_{\rm halo}-M_{\rm GCS}$ relation to DF2 gives a halo mass of $\sim 2 \times 10^{11}\Msun$ and a corresponding dark-to-stellar mass ratio of $\sim 1000$. Extrapolation of the stellar mass-halo mass relation inferred from subhalo abundance matching \citep[e.g.,][]{behroozi13, rodriguezpubela17} to lower stellar masses endows it with a halo of mass $\sim 6 \times 10^{10}\Msun$ and a corresponding dark-to-stellar mass ratio of $\sim 400$. However, the kinematics of the GCs is consistent with the galaxy being severely dark matter deficient: from the velocity dispersion of the GC system, \citet{vandokkum18a, vandokkum18c} infer a dark-to-stellar mass ratio of order unity or less inside a three-dimensional (3D) radius of $7.6\kpc$ at $90 \%$ confidence. This result has led to a spirited debate on the mass of DF2 and on whether it has a (significant) dark matter halo or not (see Section~\ref{sec:previouswork}).

An interesting constraint can be inferred from the dynamical evolution of the GCs. Since the GCs in DF2 are more luminous and therefore more massive than the average Milky Way GC, they are more susceptible to orbital decay via dynamical friction. Thus, if a mass model for the galaxy predicts very short inspiral timescales from current GC locations to the galaxy center, it is unlikely for the GCs to be observed at those locations, and the said model can be ruled out. \citet{nusser18a} explores this idea by studying the motion of a single GC in different $N$-body models for the galaxy. If DF2 is modeled as a stellar system embedded in a $10^{8}\Msun$ dark matter halo with an Einasto profile, GC orbits decay within 2-6 Gyr depending on the starting radius and the mass of the GC particle. These are varied within $(1-6)\kpc$ and $(1 -4) \times 10^{6}\Msun$ respectively. Increasing the halo mass to $10^{9}\Msun$ is also insufficient to attain inspiral timescales that exceed the age of the GCs ($\sim 10\Gyr$).

While insightful, the \citet{nusser18a} study has a number of shortcomings. First, it only considers GC orbits with an initial eccentricity of 0.5, rather than assuming a full distribution of eccentricities. Second, the GC mass range explored is on the high side as assuming a distance of 20 Mpc to DF2, the most massive GC is only about $1.5 \times 10^{6}\Msun$. Third, while inspiral timescales can't be too short, requiring them to be larger than the age of the GCs is too conservative. It is perfectly reasonable for a GC to start out on a less bound orbit and to slowly evolve over time to its present orbit. Fourth, as each simulated galaxy only has a single GC particle, GC-GC interactions are neglected, which could be significant. Fifth, \citet{nusser18a} initializes his $N$-body models for DF2 by imparting all particles (other than the GC) a speed equal to the local circular speed. As we discuss in Section~\ref{sec:summary}, this results in a very specific and unlikely orbital decay. Last but not least, given the current constraint on its halo mass, it is possible for DF2 to be a purely baryonic galaxy. Such a `baryon-only' mass model could result in slower decay of GC orbits than the `baryon-dominated' model of \citet{nusser18a} that has a $10^{8}\Msun$ halo. This is because the stellar system in DF2 has a much shallower central density profile than the Einasto profile assumed for the dark matter halo that dominates the central density in all mass models of \citet{nusser18a}.

In this paper, we investigate the dynamical evolution of DF2's GC system in the absence of dark matter. We set up a suite of 50 multi-GC $N$-body simulations matching both the projected surface brightness profile of DF2 and the projected positions and line-of-sight (los) velocities of the GCs. The GC position coordinates along the los and the velocity components perpendicular to the los are sampled from a distribution function constructed by assuming the GC system to be in equilibrium with the stellar potential. Compared to the \citet{nusser18a} study, we use an order of magnitude more particles to represent the galaxy and allow for variation in the initial orbital eccentricity of each GC consistent with its observed projected position and los velocity. Our GC particles have observationally consistent masses, and we also account for interactions among the GCs. In order to highlight the effect of GC-GC interactions, we also set up 20 simulations with only a single GC. All simulations are run forward in time for $10\Gyr$. The effect of dynamical friction on the evolution of the GCs and the response of the stellar system is studied. 

We find that on average, the GC system becomes significantly more compact over time. An example is shown in Figure~\ref{fig:image}, which compares the Hubble Space Telescope (HST) $V_{606}$-band image of DF2 (left-hand panel) to snapshots from one of our multi-GC simulations at $t=0$ (middle panel) and $t=10\Gyr$ (right-hand panel). The locations of the 10 GCs considered in this paper are indicated. Importantly, none of the GCs manage to sink all the way to the center. As we demonstrate, this is due to the fact that dynamical friction ceases to be effective in the central region of the galaxy due to the shallow inner density profile of the stellar system (`core stalling') and due to GC-GC interactions that act as an additional source of dynamical buoyancy. We, therefore, conclude that a baryon-only mass model is perfectly viable as long as some of the GCs formed somewhat further out than where we observe them today.

This paper is organized as follows. Section \ref{sec:previouswork} discusses relevant previous work on DF2 and presents observational constraints on its stellar component and its GC population. Section \ref{sec:setup} describes how the simulation suite is setup using these constraints. The results of our simulations are presented in Section \ref{sec:results}. They are summarized and discussed in Section \ref{sec:summary}. 

\section{NGC1052-DF2} 
\label{sec:previouswork}

DF2 was identified in deep, wide-field observations of the NGC 1052 group with the Dragonfly Telephoto Array \citep{abraham14,merritt16}. Using follow-up imaging data obtained with HST (see left-hand panel of Figure~\ref{fig:image}), its central surface brightness, effective radius along the major axis, S\'{e}rsic index, and axis ratio have been determined to be $\mu(V_{606},0)=24.4\ \rm mag/arcsec^2$, $R_\rme=22.6 \arcsec$, $n=0.6$ and $b/a=0.85$ respectively \citep{vandokkum18a}. 

There has been some debate on the correct distance to the galaxy. Using the surface brightness fluctuation method (SBF), \citet{vandokkum18a} determine a distance of $D=19.0 \pm 1.7\Mpc$. In an independent SBF analysis, \cite{blakeslee18} find $D=20.4 \pm 2.0\Mpc$. However, \citet{trujillo18} claim to have detected the tip of the red giant branch (TRGB) in DF2 and estimate a significantly smaller distance of $\sim 13\Mpc$. In response, \citet{vandokkum18d} show that the corresponding color-magnitude diagram is strongly influenced by blends, which can lead to an erroneous TRGB distance. Using a megamaser-TRGB-SBF distance ladder, \citet{vandokkum18d} obtain $D=18.7 \pm 1.7\Mpc$. Following \citet{vandokkum18a}, we adopt $D=20\Mpc$, resulting in $R_\rme=2.2\kpc$ along the major axis. This makes DF2 a UDG ($R_\rme > 1.5\kpc$ and $\mu(g,0)>24\ \rm mag/arcsec^2$) and also a satellite of the elliptical galaxy NGC 1052. 

Unlike other well-studied UDGs such as DF17, DF44, or VCC1287, the UDG DF2 has an unprecedented population of compact objects \citep{vandokkum18b}. There are eleven such spectroscopically confirmed objects associated with the galaxy.\footnote{see also \citet{emsellen18} who confirmed the presence of another GC in DF2.} Similar to GCs in a typical galaxy, the spatial distribution of these objects is (slightly) more extended compared to that of the smooth stellar light. Their half-number radius in projection is $\sim 3.1\kpc$, about 1.4 times larger than the projected stellar half-light radius (see Figure~\ref{fig:initial_sd_cgn}), and the outermost confirmed object is at a projected radius of $7.6\kpc$. They also resemble GCs in their compact morphologies (the objects are just resolved with HST) and colors. They are old and metal-poor with an average age of $9.3^{+1.3}_{-1.2}\Gyr$ and an average ${\rm [Fe/H]}=-1.35 \pm 0.12$, obtained from their stacked Keck spectrum. The inferred average mass-to-light ratio is $M/L_V=1.8 \pm 0.2$.

\begin{table}
    \begin{center}
    \begin{tabular}{| c | c | c | c | c | c | c |}
    \hline			
    Id & mass & $x$  & $y$  & $v_{\rm los}$  & $v_z=f v_{\rm los}$ \\
    & ($\Msun$) & ($\kpc$) & ($\kpc$) & ($\kms$) & ($\kms$) \\
    \hline  
    39 & $7.3 \times 10^5$ & -2.5 & -7.1 & $16^{+7}_{-7}$ & 11.2\\
    59 & $5.0 \times 10^5$ & 1.9 & -4.4 & $-3^{+16}_{-15}$ & -2.1\\
    71 & $5.5 \times 10^5$ & -2.4 & -1.1 & $3^{+6}_{-8}$ & 2.1\\
    73 & $1.5 \times 10^6$ & 2.1 & -0.6 & $12^{+3}_{-3}$ & 8.4\\
    77 & $9.6 \times 10^5$ & -0.4 & -0.2 & $2^{+6}_{-6}$ & 1.4\\
    85 & $6.6 \times 10^5$ & 1.4 & 0.6 & $-1^{+5}_{-6}$ & -0.7\\
    91 & $6.6 \times 10^5$ & -6.7 & 1.7 & $0^{+10}_{-10}$ & 0.0\\
    92 & $8.0 \times 10^5$ & 0.1 & 2.0 & $-13^{+6}_{-7}$ & -9.1\\
    98 & $4.2 \times 10^5$ & 0.8 & 3.6 & $-18^{+10}_{-10}$ & -12.6\\
    101 & $3.8 \times 10^5$ & -2.3 & 4.2 & $-2^{+13}_{-14}$ & -1.4\\
    \hline
    \end{tabular}
    \end{center}
    \caption{Observational constraints on the ten spectroscopically confirmed GCs of DF2 considered in this paper \citep{vandokkum18b,vandokkum18a,vandokkum18c}. Columns 1 through 5 list the Id, mass, projected spatial coordinates ($x$ and $y$) with respect to the galaxy center and los velocity with respect to the mean velocity of the GC system ($v_{\rm los}$) for each of them. $x$ and $y$ coordinates are obtained from RA and DEC measurements respectively. The los velocities are corrected for measurement error by multiplying the most probable $v_{\rm los}$ of each GC with a constant factor $f$ and listed in Column 6 (see Section~\ref{sec:setup} for  detailed discussion).}  
    \label{gc_prop} 
\end{table}

However, their luminosities are much higher than that of typical GCs. Their luminosity function has a narrow peak at $M_{V,606} \approx -9.1$, which is significantly offset from the canonical value of $M_{V}=-7.5$ \citep[e.g,][]{rejkuba12}. They are also larger in size. Their average half-light radius in projection is $\langle r_h \rangle =6.2 \pm 0.5\pc$, about a factor of 2 larger than the mean size of Milky Way GCs. Nevertheless, following \citet{vandokkum18b}, we refer to these objects as GCs and adopt $M/L_V=1.8$, which puts the mass of the most luminous GC at $1.5 \times 10^6\Msun$ and the least luminous GC at $3.8 \times 10^5\Msun$. Using the same mass-to-light ratio for the stellar system yields a stellar mass of $M_{*}=2.0 \times 10^{8}\Msun$. For comparison, the total mass in GCs is $7.6 \times 10^{6}\Msun$, about $4 \%$ of the total stellar mass.

Ten out of the eleven GCs have los velocity measurements. The observational constraints on their masses, projected positions, and los velocities are listed in Table \ref{gc_prop}. The los velocities show an unusually small spread. Using Approximate Bayesian Computation \citep[ABC;][]{beaumont02} and the square root of the variance as the measure of dispersion, \citet{vandokkum18c} obtain an intrinsic los dispersion of $\sigma^{\rm ABC}_{\rm int}=5.6^{+5.2}_{-3.8}\kms$ ($<12.4\kms$ at $90 \%$ percent confidence). Using the Tracer Mass Estimator (TME) method of \citet{watkins10}, this implies a dynamical mass of $M_{\rm dyn} < 5.2 \times 10^{8}\Msun$ inside a 3D radius of $7.6\kpc$  and a corresponding total-to-stellar mass ratio of $M_{\rm dyn}/M_{*} < 2.6$ (also within $7.6\kpc$).\footnote{The dynamical mass estimate assumes an isotropic tracer population with a power-law number density profile having a logarithmic slope of $\gamma=-1.9$ and a power law for the total potential with a logarithmic slope of $\alpha=0$.}

However, other studies have argued that the uncertainty in the velocity dispersion and the inferred halo mass could be significantly higher. Adopting a Gaussian los velocity distribution with $\sigma=\sigma_{\rm int}$ and defining the likelihood for observing a los velocity $v_{i}$ as a Gaussian with $\sigma=\sqrt{\sigma_{\rm int}^2+\delta v_{i}^2}$, where $\delta v_{i}$ is the measurement error in the $i^{\rm th}$ velocity, \citet{martin18} obtain $\sigma_{\rm int}=9.5^{+4.8}_{-3.9}\kms$ ($< 18.8\kms$ at $90 \%$ confidence). But, this estimate relies on an old velocity for one of the GCs (GC 98), quoted in \citet{vandokkum18a}. Using GC 98's revised velocity, \citet{vandokkum18c} show that the same method yields an intrinsic dispersion of $\sigma^{\rm ML}_{\rm int}=7.8^{+5.2}_{-2.2}\kms$ ($<14.6\kms$ at $90 \%$ confidence), which is consistent with $\sigma^{\rm ABC}_{\rm int}$ at the $1\sigma$ level. Using the TME method, this implies a total-to-stellar mass ratio of $M_{\rm dyn}/M_{*}<3.6$ inside $7.6\kpc$. \citet{laporte18} argue that the dispersion could be systematically underestimated due to the small number of tracers and the errors in their measured los velocities being of the same order as the dispersion. Using $10^{4}$ sets of mock velocities drawn from a Gaussian distribution of known dispersion, \citet{laporte18} find that for $\sim 10$ tracers, the median dispersion inferred from the mocks using the maximum likelihood method is about nine-tenths of the true dispersion. Taking this correction into account, they report a dispersion of $\sigma_{\rm int}=10^{+10}_{-5.5}\kms$ at $95 \%$ confidence, which again using the TME method translates to a total-to-stellar mass ratio of $M_{\rm dyn}/M_{*}<6.8$ inside $7.6\kpc$.

Using spherically symmetric Jeans models and the assumption that the GCs follow an exponential number density profile, \citet{washerman18} infer the dark matter mass within 10 kpc to be $< 1.2 \times 10^{8}\Msun$ at $90\%$ confidence, which is about three-fifths of the mass in stars. This estimate is based on the assumption that the dark matter halo of DF2 has a generalized NFW profile. Using the same methodology but adopting a more general halo profile, \citet{hayashi18} infer a total-to-stellar mass ratio of $M_{\rm dyn}/M_{*}<14.3$ inside $7.6\ \rm kpc$ at $90 \%$ confidence. However, if they adopt a power-law number density profile for the GCs, they obtain $M_{\rm dyn}/M_{*}<2.2$ at the same confidence. Finally, \citet{nusser19} uses a more general distribution function based approach to constrain the halo mass. The GCs are assumed to follow a power-law number density profile, and the dark matter halo is modeled with a tidally truncated NFW profile. The total-to-stellar mass ratio within 10 kpc is found to be $<20.5$ ($<5.9$) at $2 \sigma$ for a truncation radius of $10\kpc$ ($20\kpc$).

A more precise halo mass measurement is possible with stellar kinematics. Using an integrated spectrum of the diffuse stellar light in DF2 obtained with the Keck Cosmic Web Imager (KCWI), \citet{danieli19} find a los dispersion of $8.4^{+2.1}_{-2.1}\kms$ at $1 \sigma$. The stellar dispersion translates to a dynamical mass of $M_{\rm dyn}=1.4^{+0.7}_{-0.8} \times 10^{8}\Msun$ within the 3D circularized half-light radius of $2.7\kpc$ using the mass estimator of \citet{wolf10}. This is consistent with the total stellar mass, $M_{*}(<2.7\kpc)=1.0^{+0.2}_{-0.2} \times 10^{8}\Msun$, within the same radius \citep[see][]{vandokkum18a}. Using a different dataset obtained with the VLT/MUSE spectrograph, \citet{emsellen18} find a dispersion of $16.3^{+5}_{-5}\kms$ at $1 \sigma$ for the stellar body, which is significantly higher than the \citet{danieli19} value. While the origin of the discrepancy between the two results is not clear, \citet{danieli19} emphasize the difference in spectral resolution between the two instruments ( $\sim 12\kms$ for KCWI {\it vs.} $\sim 35-80\kms$ for VLT/MUSE). Due to a much better instrumental resolution, the higher dispersion reported by \citet{emsellen18} should have been easily detected with KCWI.

Clearly, the dark matter mass of DF2 carries a considerable amount of uncertainty. None of the existing constraints, however, can rule out the complete absence of dark matter. In this paper, we test the viability of such a `baryon-only' model by examining whether in the absence of dark matter, the observed GC population experiences rapid orbital decay due to dynamical friction. If that is the case, we may argue that the current phase-space coordinates of the GCs are highly unlikely, thereby ruling out the baryon-only hypothesis.

\section{Simulation SetUp}
\label{sec:setup}

\begin{figure*}[t]
    \centering
    \includegraphics[width=0.9\textwidth]{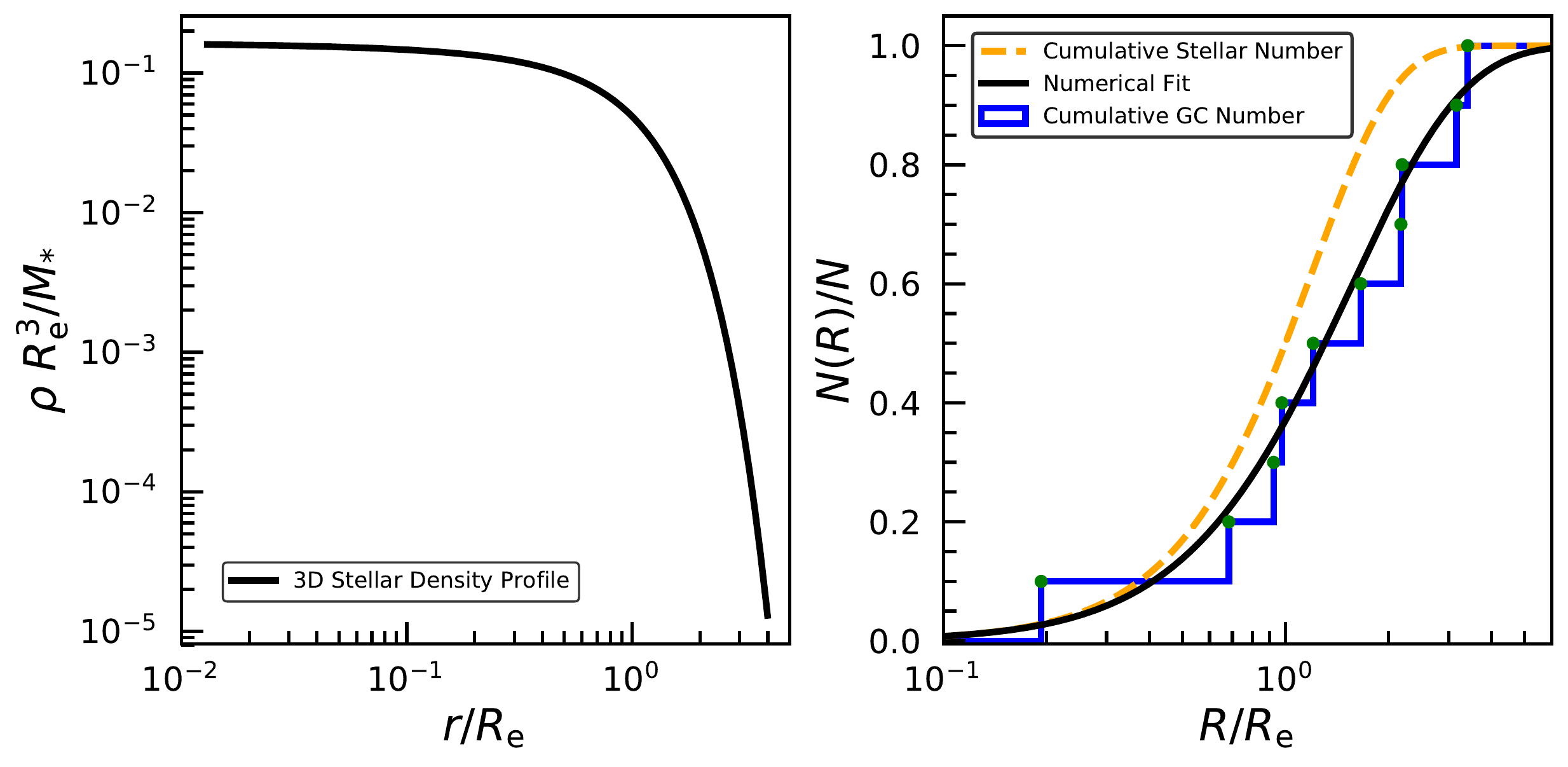}
    \caption{Left-hand panel- 3D density profile of the galaxy, which is used to generate initial conditions for the stars. It is obtained by multiplying the luminosity density profile, inferred from the observed surface brightness profile using inverse Abel transformation, with the mass-to-light ratio of DF2. Right-hand panel- observed cumulative GC number profile in projection normalized by the total number of GCs is shown by the blue histogram. Assuming a S\'{e}rsic profile for the GC number density in projection, a fit to the normalized cumulative number profile is obtained. The best fitting curve (S\'{e}rsic index = 1 and $R_{\rm half,GC}=1.3\ R_\rme$ ) is shown in black. It is used to generate initial conditions for the GC system. For comparison, the enclosed number profile of the stars in projection is shown in orange, which reveals that the spatial distribution of the GCs is slightly more extended compared to that of the stars.}
    \label{fig:initial_sd_cgn}
\end{figure*}

We model DF2 as a spherically symmetric system and compute its three-dimensional (3D) luminosity density profile, $j(r)$, by deprojecting the observed surface brightness profile (ignoring the slight non-circularity of its isophotes), $\mu(R)$, using the inverse Abel transformation,
\begin{equation}
j(r)=-\frac{1}{\pi} \int_{r}^{\infty} \frac{\rmd \mu}{\rmd R} \frac{\rmd R}{\sqrt{R^2-r^2}}\,.
\label{abel_t}
\end{equation}
The corresponding 3D density profile is given by $\rho(r)=M/L_V \ j(r)$ and is shown in the left-hand panel of Figure~\ref{fig:initial_sd_cgn} out to $4\ R_\rme$. We further assume that the galaxy is in equilibrium and that the stellar distribution function (DF) is ergodic; i.e, it depends on the phase space coordinates only through the Hamiltonian $H({\bf r},{\bf v})$. This allows the DF to be computed using the Eddington equation,
\begin{equation}
f(\epsilon)=\frac{1}{\sqrt{8} \pi^2} \frac{\rmd}{\rmd \epsilon} \int_{0}^{\epsilon} \frac{\rmd \psi}{\sqrt{\epsilon-\psi}} \frac{\rmd \rho}{\rmd \psi}\,.
\label{eddington_f}
\end{equation}
Here $\psi = -\phi$, $\phi$ being the stellar potential corresponding to the density, $\rho$, and $\epsilon=-H=\psi-\frac{1}{2}{v^2}$. The DF is then used to sample positions and velocities of the particles representing the stellar system. 

To generate initial conditions for the GCs, we parameterize the projected GC number density with a S\'{e}rsic profile,
\begin{equation}
\Sigma (R) \propto {\rm exp} \left[ -b \left( \frac{R}{R_{\rm half,GC}} \right)^{1/\alpha} \right]\,.
\label{gc_pden}
\end{equation}
Here $\alpha$ is the S\'{e}rsic index; $R_{\rm half,GC}$ is the GC half-number radius in projection, and $b$ satisfies $\gamma (2\alpha;b) = \Gamma(2\alpha)/2$ where $\Gamma$ and $\gamma$ are the Gamma function and lower incomplete Gamma function respectively. The normalized cumulative GC number profile in projection is given by
\begin{equation}
\frac{N(R)}{N}= \frac{\int_{0}^{R} \Sigma(R) R \, \rmd R}{\int_{0}^{\infty} \Sigma(R) R \, \rmd R}\,.
\label{gc_cn}
\end{equation}

The blue histogram in the right-hand panel of Figure~\ref{fig:initial_sd_cgn} shows the observed cumulative GC number profile in projection normalized by the total number of GCs. It is fitted with Equation~\ref{gc_cn} using the least squares method, and the best fitting curve is shown in black. The best fit is obtained for $\alpha=1$ and $R_{\rm half,GC}=1.3\ R_\rme$, which are the parameters we adopt throughout. The corresponding 3D number density profile of the GCs, $n(r)$, is obtained from $\Sigma(R)$ using the inverse Abel transformation under the assumption of spherical symmetry. Under the additional assumption that the GCs are isotropic, mass-less tracers of the stellar potential, we use the Eddington equation to compute their distribution function, which requires replacing $\rho$ in Equation~\ref{eddington_f} with $n$. 

In order to draw positions and velocities for the GCs, we proceed as follows. We assume that the los velocity of DF2's stellar body is equal to the mean velocity of the ten GCs \citep[see][]{emsellen18, danieli19}. However, the los velocity of each GC with respect to that mean needs to be corrected for measurement error. After all, the observed los velocity dispersion, $\sigma_{\rm obs}=10.1\kms$, is larger than the intrinsic los velocity dispersion, $\sigma_{\rm int}$. We can estimate the typical measurement error using $\sigma_{\rm err}=\sqrt{\sigma_{\rm obs}^2-\sigma_{\rm int}^2}$ and the estimates for $\sigma_{\rm int}$ obtained by \citet{vandokkum18c}. The best estimate for $\sigma_{\rm int}=5.6\ \rm km\ s^{-1}$ (ABC) or $7.8\ \rm km\ s^{-1}$ (ML). Hence, $\sigma^{\rm ABC}_{\rm err}=8.4\kms$ or $\sigma^{\rm ML}_{\rm err}= 6.4\kms$. Taking the mean of the two, we get $\sigma_{\rm err}=7.4\kms$, which we use to correct the los velocities, $v_{\rm los}$, of the GCs with respect to their mean by multiplying them with a correction factor, $f \equiv \sqrt{\sigma_{\rm obs}^2 - \sigma_{\rm err}^2} / \sigma_{\rm obs} \approx 0.7$. The corrected los velocities, $v_z$, are listed in column~6 of Table~\ref{gc_prop}.

The projected positions ($x,y$) and corrected los velocities ($v_z$) are used as constraints to make 50 random realizations for the GC system by sampling the velocity components perpendicular to the los ($v_x$,$v_y$) and the positions along the los ($z$). Given $x$, $y$, and $v_z$ for a GC, $z$, $v_x$, and $v_y$ are sampled from $P(z|x,y,v_z)$, $P(v_x|x,y,v_z,z)$, and $P(v_y|x,y,v_z,z,v_x)$ respectively. Here $P(z|x,y,v_z)$, $P(v_x|x,y,v_z,z)$, and $P(v_y|x,y,v_z,z,v_x)$ are given by 
\begin{equation}
P(z|x,y,v_z)=\frac{\int_{v_{x}^{-}}^{v_{x}^{+}} \int_{v_{y}^{-}}^{v_{y}^{+}} f({\bf r},{\bf v}) \ \rmd v_y \ \rmd v_x}{\int_{z^{-}}^{z^{+}} \int_{v_{x}^{-}}^{v_{x}^{+}} \int_{v_{y}^{-}}^{v_{y}^{+}} f({\bf r},{\bf v}) \ \rmd v_y \ \rmd v_x \ \rmd z} \,,
\end{equation}
\begin{align}
&P(v_x|x,y,v_z,z)= \frac{\int_{v_{y}^{-}}^{v_{y}^{+}} f({\bf r},{\bf v})\ \rmd v_y}{ \int_{v_{x}^{-}}^{v_{x}^{+}} \int_{v_{y}^{-}}^{v_{y}^{+}} f({\bf r},{\bf v})\ \rmd v_y \ \rmd v_x} \,,\\
&P(v_y|x,y,v_z,z,v_x)= \frac{f({\bf r},{\bf v})}{\int_{v_{y}^{-}}^{v_{y}^{+}} f({\bf r}, {\bf v})\ \rmd v_y} \,, 
\end{align}
where f is the GC distribution function, ${\bf r}=(x,y,z)$ and ${\bf v}=(v_{x},v_{y},v_{z})$. The integration limits are derived from the requirement that $v = \sqrt{v^2_x + v^2_y + v^2_z} \leq v_{\rm esc}(r)$ and are given by $v^{\pm}_{y}=\pm \sqrt{v^2_{\rm esc}(r)-v_z^2-v_x^2}$, $v^{\pm}_{x}= \pm \sqrt{v^2_{\rm esc}(r)-v_z^2}$, and $z^{\pm}=\pm \sqrt{r^2_{\rm max}-x^2-y^2}$. Here $v_{\rm esc}(r)$ is the escape velocity at the 3D radius, $r$, and $r_{\rm max}$ is the radius where $v_{\rm esc}=|v_z|$. 

\begin{figure}
    \centering
    \includegraphics[width=0.4\textwidth]{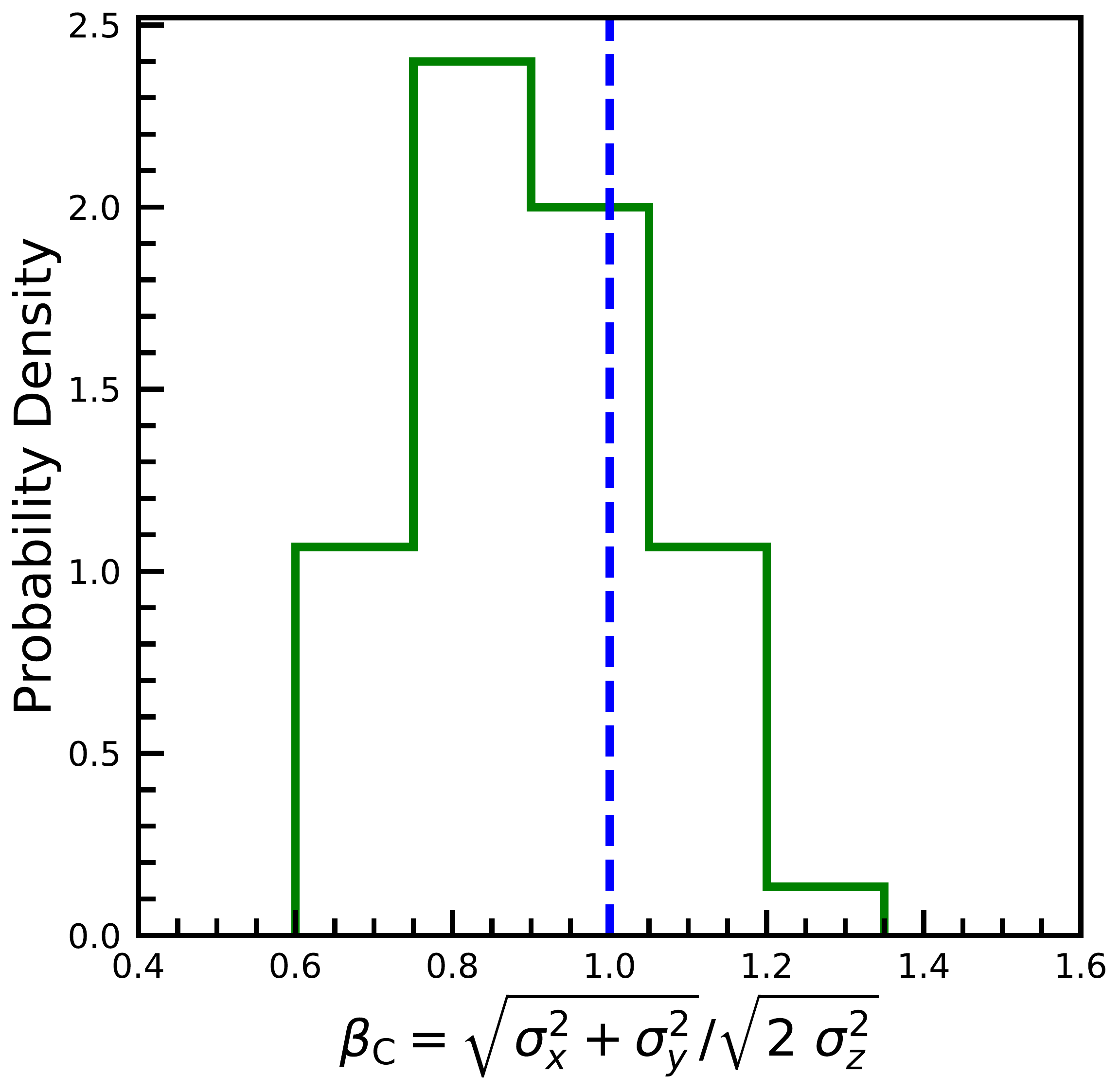}
    \caption{Probability distribution of $\beta_{\rmC}$, a parameter that measures the velocity anisotropy of the GC system, obtained from 50 random realizations at $t=0$. We find that $\beta_\rmC = 0.88^{+0.26}_{-0.21}$ (where we quote the median, $5^{\rm th}$ and $95^{ \rm th}$ percentiles), which is perfectly consistent with an isotropic distribution ($\beta_\rmC = 1$), indicated by the blue, dashed vertical line. Hence, the observed kinematics of the GCs is consistent with what is expected for an isotropic tracer population in equilibrium with the (spherically symmetric) stellar potential.}
    \label{fig:anisotropy}
\end{figure}

Figure~\ref{fig:anisotropy} shows the probability distribution of the Cartesian anisotropy parameter,
\begin{equation}
\beta_\rmC \equiv \sqrt{ \frac{\sigma^{2}_{x} + \sigma^{2}_{y}}{2\ \sigma^{2}_{z}}} \, ,
\end{equation}
obtained for the 50 random realizations thus constructed. Here  $\sigma_{x}$, $\sigma_{y}$, and $\sigma_{z}$ are the rms dispersions of $v_{x}$, $v_{y}$, and $v_{z}$ respectively. While $\sigma_z = 7.1\kms$ for each realization, by construction, $\sigma_{x}$ and $\sigma_{y}$ vary from one realization to another. Within the $5^{\rm th}$-$95^{ \rm th}$ percentile, we find that $0.67 < \beta_\rmC < 1.14$, which is in excellent agreement with isotropy ($\beta_\rmC=1$), indicated by the blue, dashed vertical line. This indicates that the corrected los velocities of the GCs are consistent with our assumed mass model and with the assumed spherical, isotropic nature of the GC system. For instance, if $\beta_\rmC$ would have been significantly larger than unity, it would have implied that we happen to observe the GC system at a peculiar, unlikely moment in time when their 3D distribution is reminiscent of a face-on disk structure. Alternatively, if $\beta_\rmC$ had been significantly smaller than unity, it would have signaled the need for additional (probably dark) matter. Hence, we conclude that the observed kinematics of the GCs is consistent with what is expected for an isotropic tracer population in equilibrium with the (spherically symmetric) stellar system.

We run a total of 50 multi-GC $N$-body simulations, each comprising of $10^{6}$ star particles (hereafter stars) and 10 GCs. All 50 simulations use the same set of initial conditions for the stars, each having a mass of $200 \Msun$, but a different realization for the initial phase-space coordinates of the GCs. In addition, we run a set of 20 single-GC simulations, also having $10^{6}$ stars each but only one GC (GC 92, the third most massive GC in DF2); these simulations only differ in the initial los position and tangential velocity of GC 92 and are used for comparison with the multi-GC simulations in order to gauge the impact of GC-GC interactions. Finally, we also run a single simulation with only stars (no GCs). Both the stars and the GCs are represented as Plummer spheres with a softening length of $10\pc$, and all simulations are run for 10 Gyr using a Barnes-Hut octree code \citep{barnes86} with an opening-angle of $\theta=0.7$ and a leapfrog integration scheme with a fixed time step of $8 \times 10^{5}\yr$. The total energy among all runs is conserved to better than $1\%$ over 10 Gyr. As detailed in Appendices~\ref{force_soft} and~\ref{time_step}, these choices for the softening length and the time step are adequate for properly resolving the dynamical friction experienced by the GCs.

\section{Results}
\label{sec:results}

\subsection{Realizations with a single GC}
\label{sec:singleGC}

\begin{figure*}
    \centering
    \includegraphics[width=\textwidth]{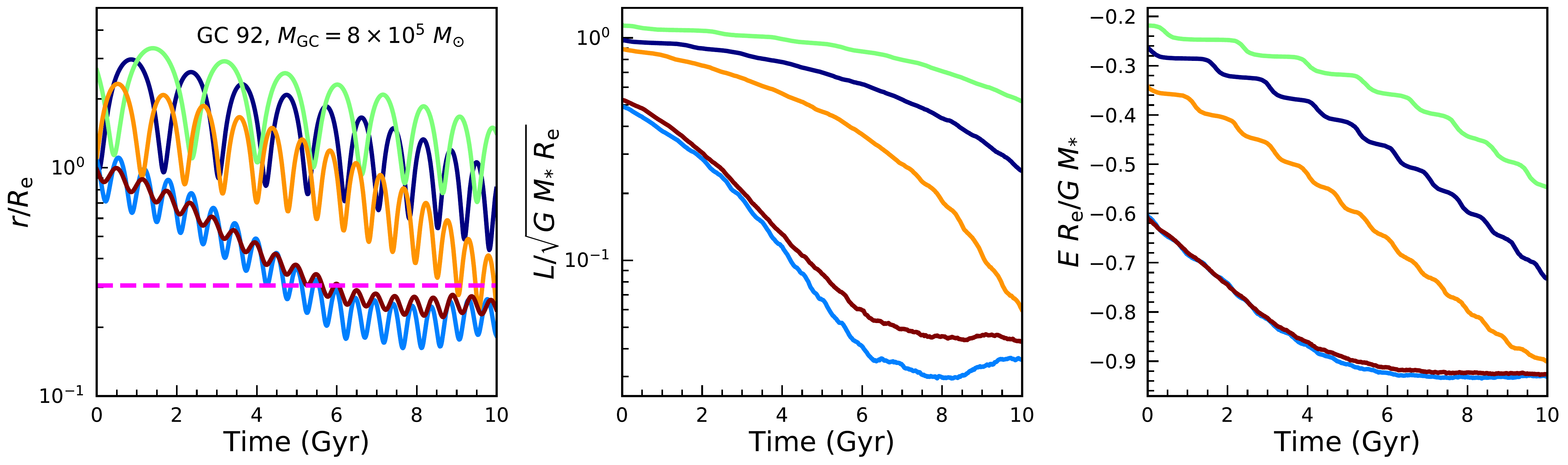}
    \caption{Evolution of the 3D radius (left-hand panel), specific angular momentum (middle panel), and specific energy (right-hand panel) of GC 92 in a random subset of 5 of the 20 simulations that are run with a single GC. The center of mass of all stellar particles is chosen as the frame of reference. There is a large realization-to-realization variance in the evolution of the GC. The amount of orbital decay experienced by the GC is correlated with its initial energy - it sinks towards the galaxy center more in those realizations where it is initially more bound. However, irrespective of initial conditions, when it reaches a radius close to $\sim 0.3\ R_\rme$, it stops sinking further. This phenomenon is known as core stalling. The dashed, magenta line in the left-hand panel indicates $r_{*}$ for GC 92, which is defined as the root of Equation~\ref{rchar}. As predicted by \citet{kaur18}, the onset of core stalling occurs close to this radius.}
    \label{fig:gc_single}
\end{figure*}

\begin{figure*}
    \centering
    \includegraphics[width=0.9\textwidth]{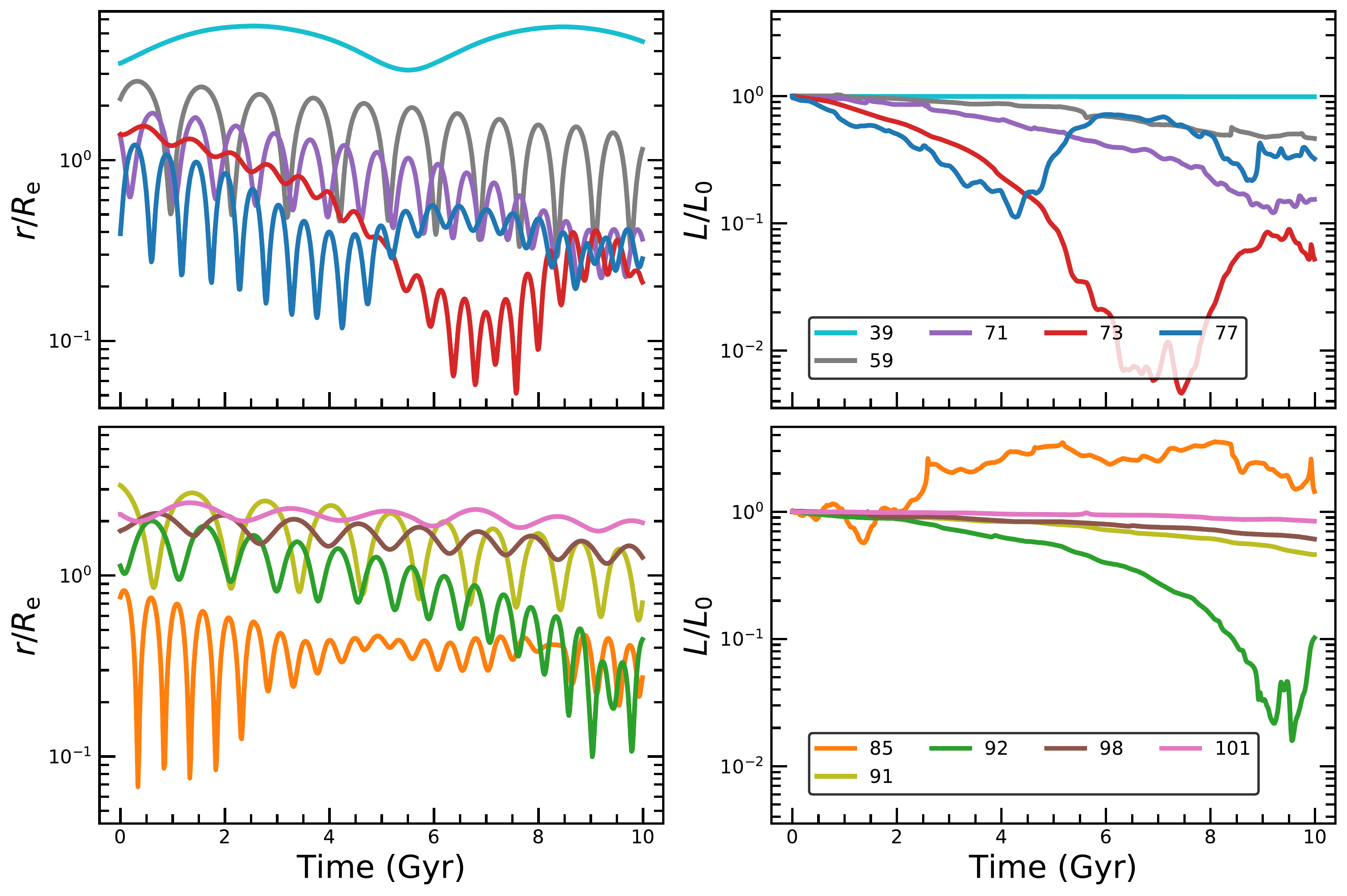}
    \caption{Evolution of the 3D radius (left-hand panels) and the orbital angular momentum (right-hand panels) of each GC in one randomly selected multi-GC simulation out of the 50. The GCs are divided into two groups of five (top and bottom panels) for clarity. While the 3D radii are measured with respect to the center of mass of all stars, the angular momenta are measured with respect to the center of mass of the total system (stars+GCs). When GC-GC interactions dominate over GC-star interactions, instead of steadily losing angular momentum to the stars, the GCs start exchanging angular momentum among themselves. In this particular realization of the GC system, the evolution of GCs 71, 73, 77, 85 and 92 is significantly impacted by such interactions. In addition to the reduced efficiency of dynamical friction in the stellar core, GC-GC scattering keeps the GCs afloat, preventing them from sinking to the center of the galaxy.}
    \label{fig:sample_realization}
\end{figure*}

\begin{figure*}
    \centering
    \includegraphics[width=0.9\textwidth]{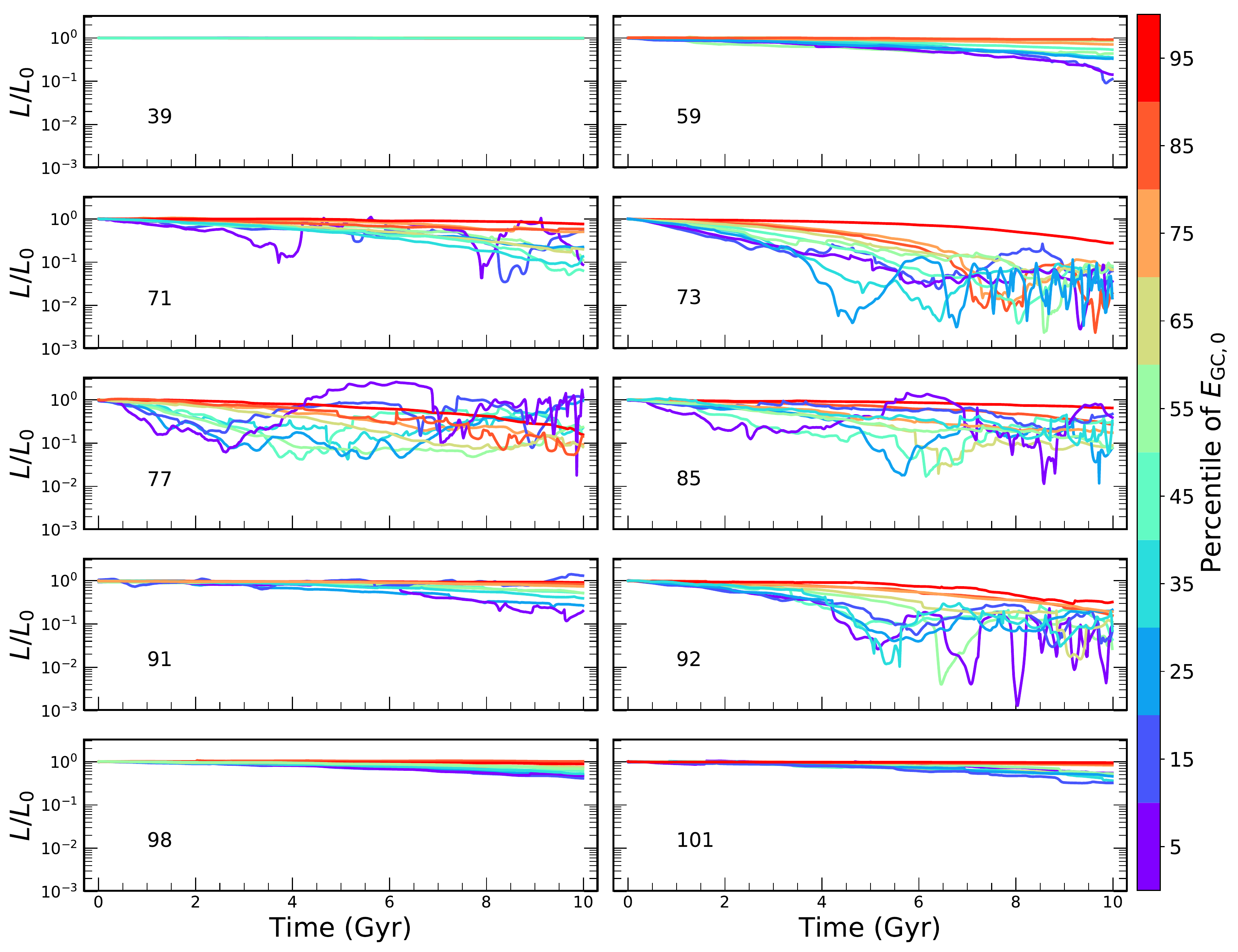}
    \caption{Evolution of the orbital angular momentum of each GC defined with respect to the center of mass of the total system (stars+GCs). For a particular GC, all 50 multi-GC simulations are rank ordered by the energy of the GC in that simulation at $t=0$ and the evolution in the simulations located at every $10^{\rm th}$ percentile starting from the $5^{\rm th}$ to the $95^{\rm th}$ is shown. Each GC exhibits a substantial amount of realization-to-realization variance (except GC 39 which is consistent with no evolution in every realization). GC-GC interactions have a stronger impact on the evolution of the five innermost GCs in projection at $t=0$ (71, 73, 77, 85 and 92). Compared to the outer GCs (39, 59, 91, 98 and 101), these globulars are also more affected by dynamical friction.}
    \label{fig:gc_momentum_individual}
\end{figure*}

\begin{figure*}
    \centering
    \includegraphics[width=0.9\textwidth]{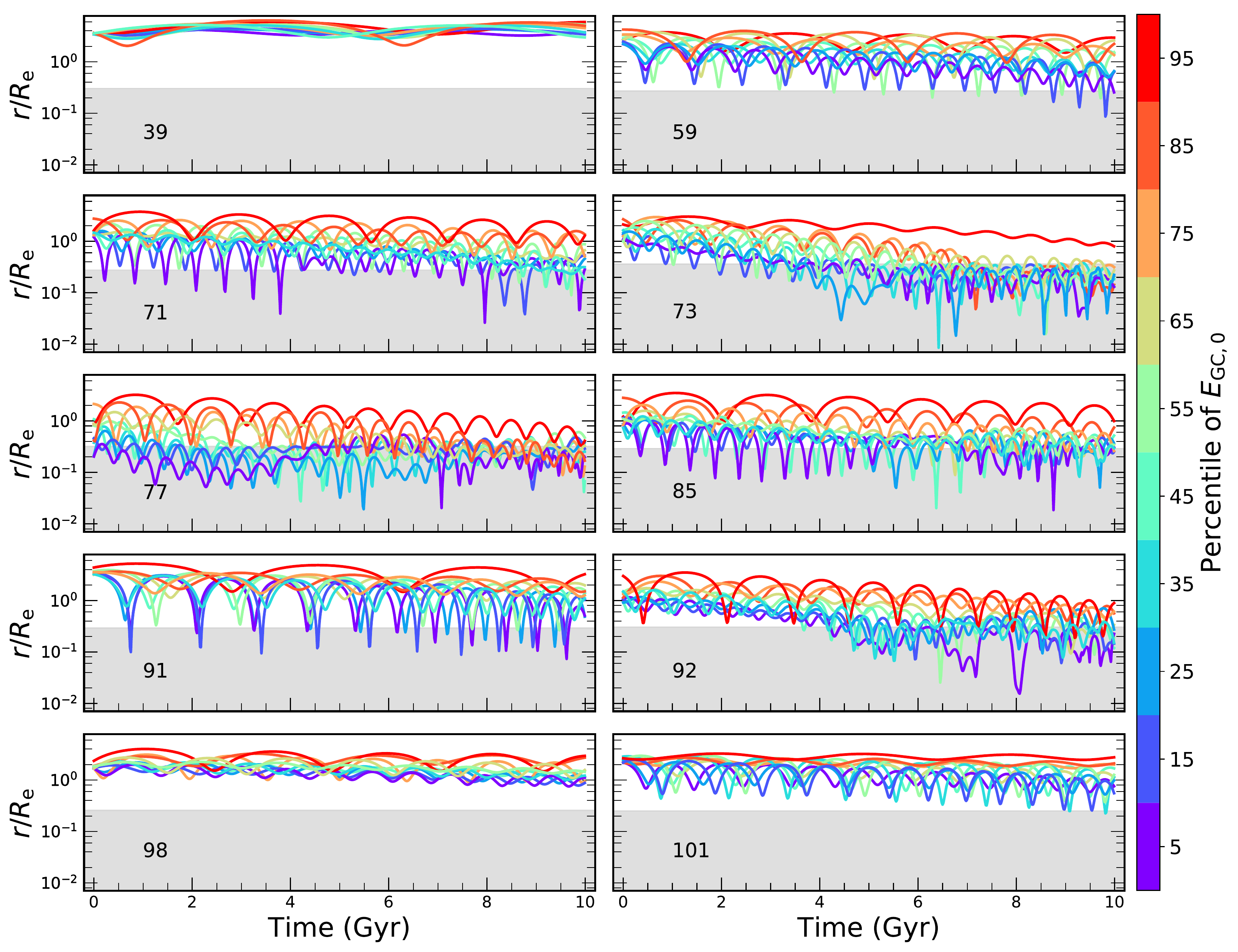}
    \caption{Evolution of the 3D radius of each GC defined with respect to the center of mass of all stars. For a particular GC, all 50 multi-GC simulations are rank ordered by the energy of the GC in that simulation at $t=0$ and the evolution in the simulations located at every $10^{\rm th}$ percentile starting from the $5^{\rm th}$ to the $95^{\rm th}$ is shown. For each GC, the region $r<r_{*}$ is shaded in grey, where $r_{*}$ is defined as the root of Equation~\ref{rchar}. Note that if GC-GC interactions were absent, a GC on a circular orbit would have stalled at a radius close to $r_{*}$ (see Section~\ref{sec:corestalling}). It could have ventured into the shaded region but only if it were on a fairly eccentric orbit. However, when multiple GCs are present, GC-GC interactions can also scatter a GC into the shaded region; a similar interaction at a later time will typically scatter it back out to a larger radius.}
    \label{fig:gc_radius_individual}
\end{figure*}

We begin by describing the results obtained from our suite of single-GC simulations involving GC 92. In Figure \ref{fig:gc_single}, we show the evolution of the 3D radius (left-hand panel), specific angular momentum (middle panel) and specific energy (right-hand panel) of the GC particle over 10 Gyr in a random subset of 5 of the 20 simulations. The center of mass of all stellar particles is chosen as the frame of reference. In each realization, dynamical friction causes the GC to move in, gradually losing energy and angular momentum. The amount of orbital decay, however, varies from one realization to another. The GC sinks towards the galaxy center more in those realizations where it is initially more bound. Interestingly, in the two cases where the GC is most bound initially (depicted by the brown and light-blue curves), the orbital decay ceases after about 6 Gyr, when the GC has reached a radius close to $\sim 0.3\ R_\rme$. This phenomenon is known as core stalling: the demise of dynamical friction in the central region of a galaxy (or halo) with a very shallow density profile \citep[e.g.,][]{hernandez98, read06, inoue09, inoue11, petts15, petts16, kaur18}. Given that the stellar mass distribution of DF2 has such a central `core', roughly inside $\sim 0.3\ R_\rme$ (see Figure~\ref{fig:initial_sd_cgn}), core stalling is expected to play an important role in regulating the rate at which its population of GCs sinks towards its center. The next subsection therefore briefly discusses this phenomenon.

\subsection{Core stalling}
\label{sec:corestalling}

Chandrasekhar's classical formulation of dynamical friction \citep[][]{chandrasekhar43} is based on the assumption that the frictional drag arises from the combined effect of uncorrelated two-body interactions between the subject mass and individual background particles. This yields the well known expression for the dynamical friction force:
\begin{equation}\label{eqn_chandra_df}
F_{\rm df}=M_{S} \frac{\rmd \bf{v}_{S}}{\rmd t}=-4 \pi \left( \frac{GM_{S}}{v_{S}} \right)^2 {\rm ln} \Lambda\ \rho(<v_{S})\ \frac{\bf v_{S}}{v_{S}} \,. 
\end{equation}
Here $M_{S}$ is the subject mass, $\bf{v}_{S}$ is its velocity, $\rho(<v_{S})$ is the density of the background particles that are moving slower than the subject mass, and ${\rm ln} \Lambda ={\rm ln} \left( b_{\rm max}/b_{\rm min} \right)$ is the Coulomb logarithm, which is introduced to control the  range of impact parameters relevant for the friction force. In particular, the maximum impact parameter, $b_{\rm max}$, is typically assumed to be of the order of the size of the host system, while $b_{\rm min}$ is typically set equal to the impact parameter for which a two-body encounter results in an angular deflection of $90^{\circ}$. 

Chandrasekhar's derivation is based on the assumption of an infinite and homogeneous `sea' of background particles, which is not a realistic description of galaxies. \citet[][hereafter TW84]{tremaine84} and \citet{weinberg86, weinberg89} developed a more realistic perturbative theory of dynamical friction that is applicable to spherical systems. They demonstrate that the torque responsible for dynamical friction arises from background particles that are close to resonance with the perturber (i.e., the subject mass). The crucial role played by resonances gives insight into the origin of core stalling. For a subject mass moving on a circular orbit, using the TW84 method, \citet{kaur18} show that in a cored density profile, the number and strength of resonances progressively decrease as the perturber moves towards the galaxy center. It is argued that the subject mass will stall at a characteristic radius, $r_{*}$, defined as the root of
\begin{equation}\label{rchar}
  \Omega(r) = \sqrt{\frac{4}{3} \, \pi \, G \, \rho_0}\,.
\end{equation}
Here $\Omega(r) = v_\rmc(r)/r$ is the circular frequency of the subject mass, and $\rho_0$ is the central density of the background system. Inside $r_{*}$, many of the strong, low-order resonances disappear, and dynamical friction becomes ineffective. The dashed, magenta line in the left-hand panel of Figure~\ref{fig:gc_single} indicates $r_{*}$ for GC 92. As is evident, the onset of core stalling indeed occurs close to this radius.

\subsection{Realizations with Multiple GCs}
\label{sec:multipleGC}

Thus far, we have only focused on the orbital evolution of a single GC as it orbits the cored stellar mass distribution of DF2. However, DF2 has multiple GCs, which potentially allows for a much richer dynamics. In particular, since the self-gravity among the various GCs is not negligible (as their total mass is about $4 \%$ of that of the stellar body), we expect GC-GC interactions to play a significant role. In addition, each GC not only feels the impact of its own response density but also that from all other GCs. Since each response density has a different pattern speed (due to the different orbital frequencies of the different GCs), the total potential in which the GCs orbit has a time-variability to it that causes a diffusion of the orbital energies of the GCs, similar to what happens in the case of violent relaxation. And since violent relaxation causes mass-independent mixing, it tends to undo the effects of dynamical friction, which instead results in mass segregation. Hence, the presence of multiple GCs may have a significant impact on the overall efficiency of dynamical friction compared to the single GC case. In order to investigate this in detail, we run a suite of 50 simulations in which we follow all 10 GCs in DF2 simultaneously.

Figure \ref{fig:sample_realization} shows the evolution of the individual GCs in one of these multi-GC simulations. The GCs are divided into two groups of five (top and bottom panels) for clarity. The left and right-hand panels depict the evolution of the 3D radius and the orbital angular momentum of each GC respectively. While the 3D radii are measured with respect to the center of mass of all stars, the orbital angular momenta are measured with respect to the center of mass of the total system (stars+GCs). That way the changes in angular momenta are only due to inertial forces. 

We find that the evolution of some of the GCs is substantially affected by interactions with other GCs. When such interactions dominate over GC-star interactions, instead of steadily losing angular momentum to the stars, the GCs start exchanging angular momentum among themselves, and angular momentum gained from one or more GCs can compensate for the angular momentum lost to the stars. Therefore, in addition to the reduced efficiency of dynamical friction in the stellar core, GC-GC scattering prevents the GCs from sinking to the center of the galaxy. In the realization shown in Figure \ref{fig:sample_realization}, the evolution of GCs 71, 73, 77, 85 and 92 is significantly impacted by such interactions. The effect is most pronounced for GC 85, which ends up with more angular momentum than what it started with.

\subsubsection{Effect on Individual GCs}
\label{sec:individualGCs}

Having highlighted the importance of GC-GC interactions using one random realization of the GC system, we now focus on the variance in the evolution of the GCs from one realization to another. In Figures \ref{fig:gc_momentum_individual} and \ref{fig:gc_radius_individual}, we show the evolution in the 3D radius and the orbital angular momentum of each GC respectively over 10 Gyr. For a particular GC, all 50 multi-GC simulations are rank ordered by the energy of the GC in that simulation at t=0 ($E_{\rm GC,0}$) and the evolution in those located at every $10^{\rm th}$ percentile starting from the $5^{\rm th}$ to the $95^{\rm th}$ is shown. Except GC 39, which experiences no significant amount of dynamical friction in any of the realizations, the other GCs exhibit a substantial amount of realization-to-realization variance. 

During the initial stages of evolution, the angular momentum lost by a GC in a particular realization is correlated with $E_{\rm GC,0}$. A GC sinks towards the galaxy center more in those realizations where it is initially more bound. However, as the GCs move in, GC-GC interactions start to play a significant role. Such interactions allow the GCs to exchange energy and angular momentum among themselves. As a result, the correlation between $L/L_0$ and $E_{\rm GC,0}$ is partially washed out during the later stages of evolution. The impact of GC-GC interactions is more pronounced for the five GCs (71, 73, 77, 85 and 92) that at $t=0$ have the smallest projected radii. This is evident from the large oscillations in their orbital angular momenta, particularly in the realizations corresponding to lower percentiles of $E_{\rm GC,0}$.

\begin{figure}
    \centering
    \includegraphics[width=0.4\textwidth]{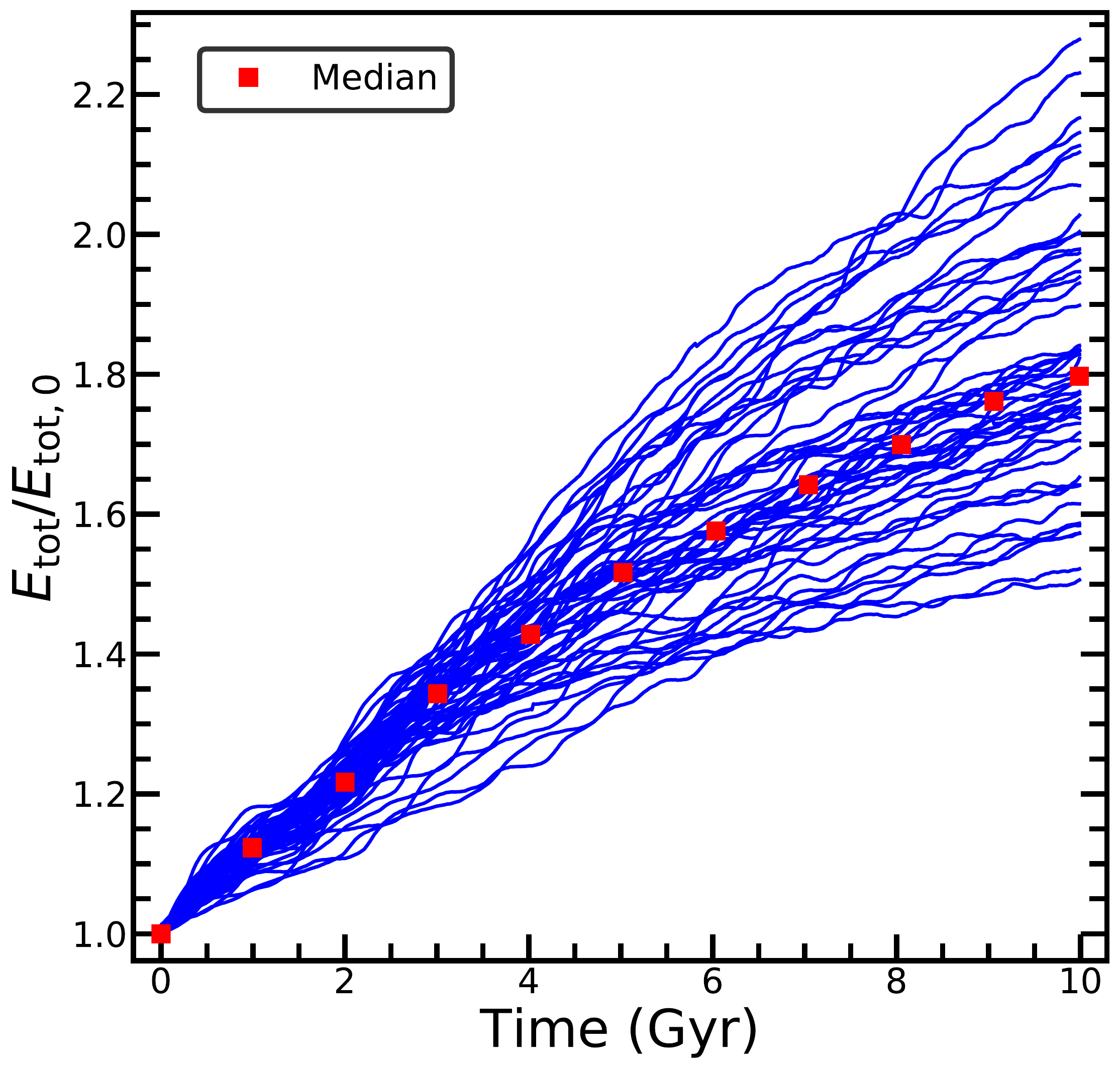}
    \caption{Evolution of the total energy of the GC system ($E_{\rm tot}$) with respect to its initial energy at $t=0$ ($E_{\rm tot,0}$) over 10 Gyr for each of the 50 multi-GC simulations. Median values of $\mathscr{E}=E_{\rm tot}/E_{\rm tot,0}$ over all 50 simulations are indicated with red squares at separations of 1 Gyr. Over time, because of dynamical friction, the GC system loses energy to the stars, becoming more bound. On average, after 10 Gyr, the binding energy of the GC system increases by about 80 percent but with a large realization-to-realization variance.}
    \label{fig:gc_system_energy_change}
\end{figure}

\begin{figure*}
    \centering
    \includegraphics[width=0.9\textwidth]{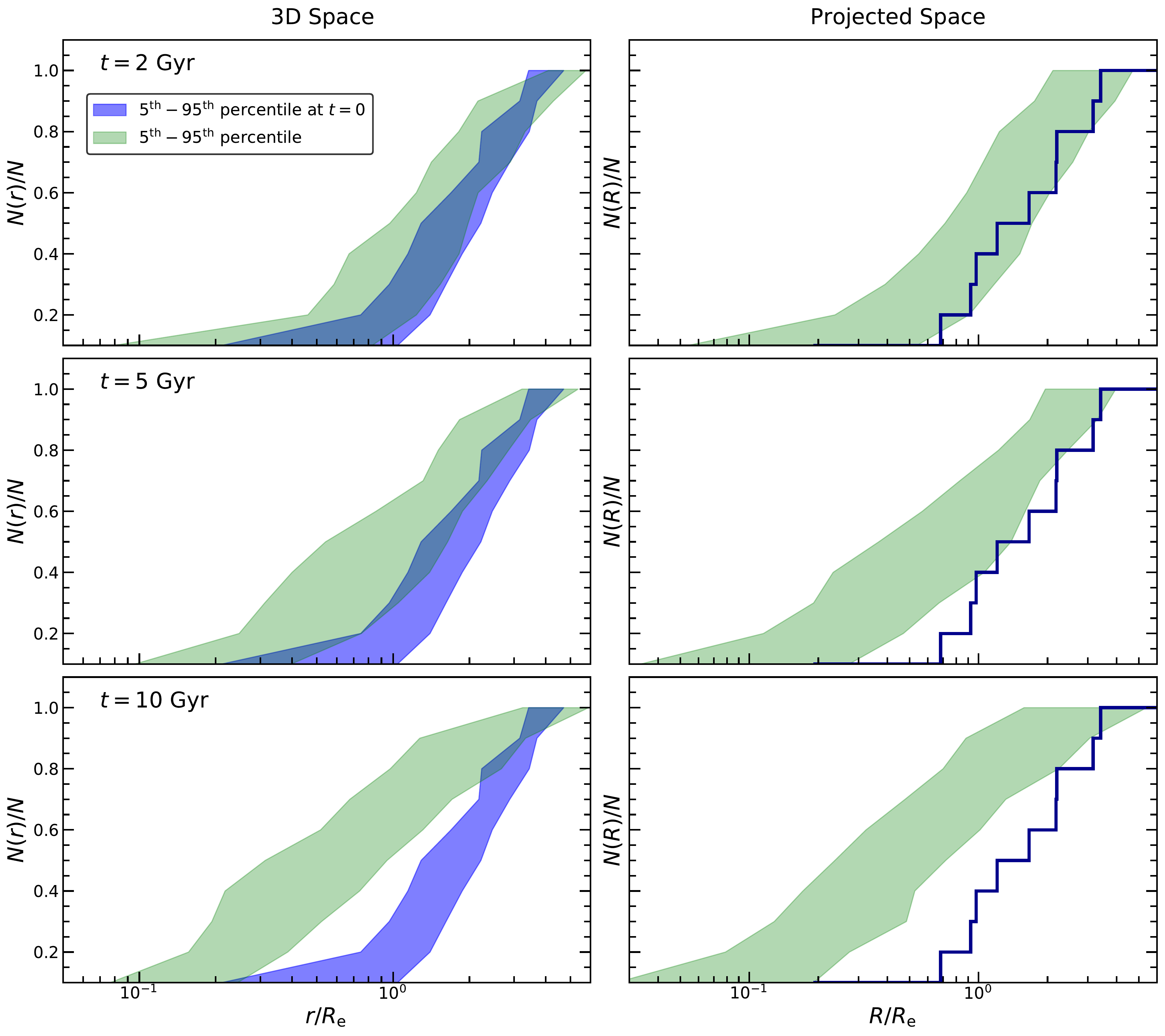}
    \caption{Left-hand panels show the variance in the 3D cumulative GC number profile ($5^{\rm th}$-$95^{\rm th}$ percentiles) at $t=0$ in blue. The variance in the evolved profile after 2, 5, and 10 Gyr is shown in green. The blue histograms in the right-hand panels depict the cumulative GC number profile in projection at $t=0$, which is same for every realization by construction. The variance in it after 2, 5, and 10 Gyr of evolution is shown in green. During the first 2 Gyr, the profiles evolve little, and the cumulative GC number profile in projection after 2 Gyr is consistent with that at $t=0$ within the realization-to-realization variance. The profiles evolve more over 5 Gyr and are significantly different after 10 Gyr.}
    \label{fig:gc_cumulative_number_profile}
\end{figure*}

\begin{figure*}
    \centering
    \includegraphics[width=0.9\textwidth]{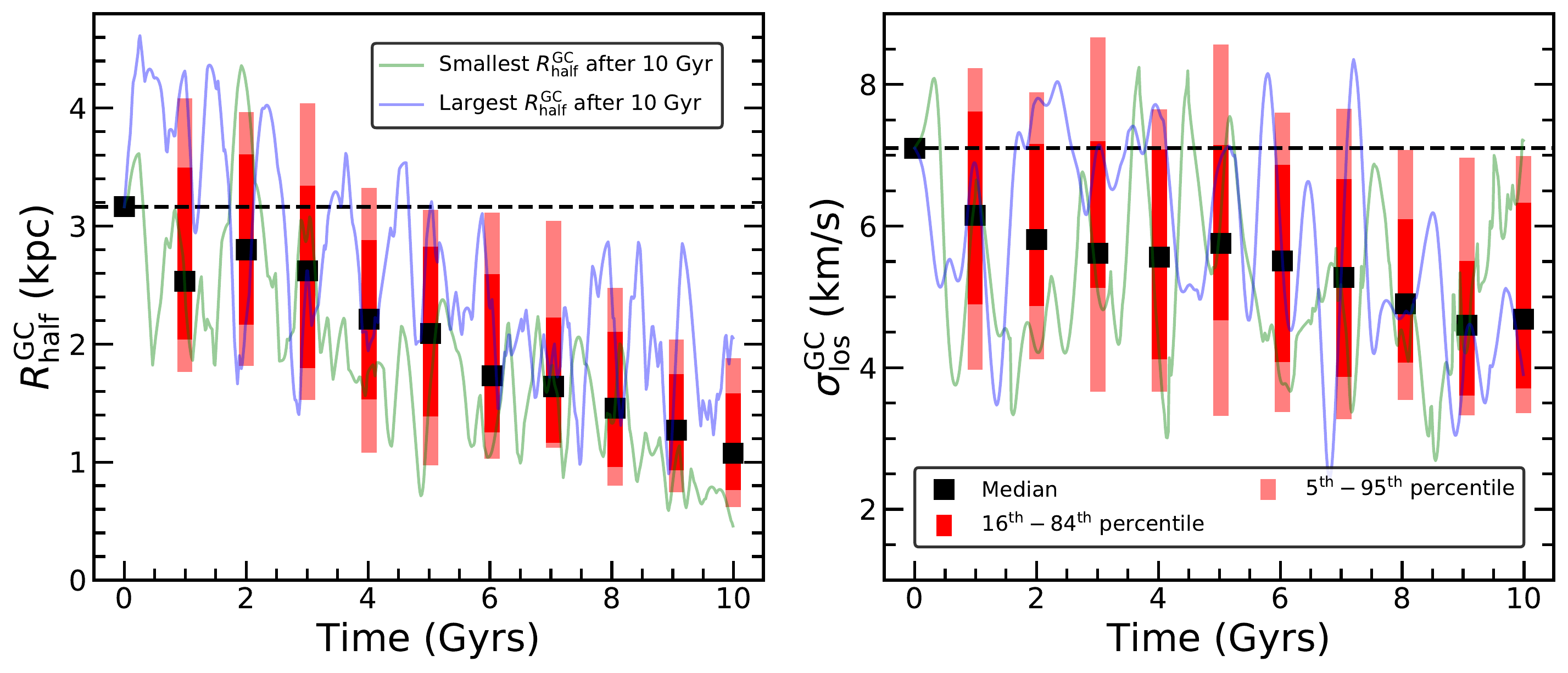}
    \caption{Left and right-hand panels show the evolution in the projected half-number radius ($R_{\rm half}^{\rm GC}$) and los dispersion ($\sigma_{\rm los}^{\rm GC}$) of the GC system over 10 Gyr respectively. $R^{\rm GC}_{\rm half}$ is defined here as the mean of the projected radius of the $5^{\rm th}$ and the $6^{\rm th}$ innermost GC and $\sigma_{\rm los}^{\rm GC}$ is the rms spread in the los velocities of the GCs. The medians obtained from the 50 multi-GC simulations are shown in black at separations of 1 Gyr. Over 10 Gyr, the median $R_{\rm half}^{\rm GC}$ shrinks by $65 \%$, while the median $\sigma_{\rm los}^{\rm GC}$ decreases by $34 \%$. However, there is also a large realization-to-realization variance, which is highlighted by indicating the $16^{\rm th}$-$84^{\rm th}$ percentiles in deep red and $5^{\rm th}$-$95^{\rm th}$ percentiles in light red respectively. Finally, the blue and green curves depict the evolution of $R_{\rm half}^{\rm GC}$ and $\sigma_{\rm los}^{\rm GC}$ for two specific realizations: those which have the largest and the smallest $R_{\rm half}^{\rm GC}$ after 10 Gyr of evolution respectively.}
    \label{fig:gc_rhalf_los_dispersion}
\end{figure*}

Compared to the outer GCs (39, 59, 91, 98 and 101), the five innermost GCs in projection at $t=0$ are also the ones which are more affected by dynamical friction. They lose more angular momentum to the stars compared to the outer GCs when realizations that are minimally affected by GC-GC interactions and located at the same percentile of $E_{\rm GC,0}$ are compared. For example, in the realizations located at the $95^{\rm th}$ percentile of $E_{\rm GC,0}$, GCs 73 and 77 lose about $70 \%$ and $90 \%$ of their initial angular momentum respectively after 10 Gyr of evolution. At the same percentile, GCs 98 and 101 experience zero evolution in their orbital angular momentum.

Due to the reduced efficiency of dynamical friction in the stellar core and due to GC-GC interactions that prevent the GCs from monotonically losing angular momentum to the stars, all GCs remain buoyant and fail to sink all the way to the center of the galaxy. In Figure~\ref{fig:gc_radius_individual}, the region $r<r_{*}$ is shaded in grey for each GC. Here $r_{*}$ is the distance from the galaxy center where core stalling is expected to take effect for the GC in question (see Equation~\ref{rchar}). A GC can venture into the shaded region when on fairly eccentric orbits or when interactions with other GCs scatter it onto more bound orbits. In the latter case, a similar interaction at some later time will typically scatter it back out to a larger radius.

\subsubsection{Implications for the GC System}
\label{sec:implications}

\begin{figure*}
    \centering
    \includegraphics[width=0.9\textwidth]{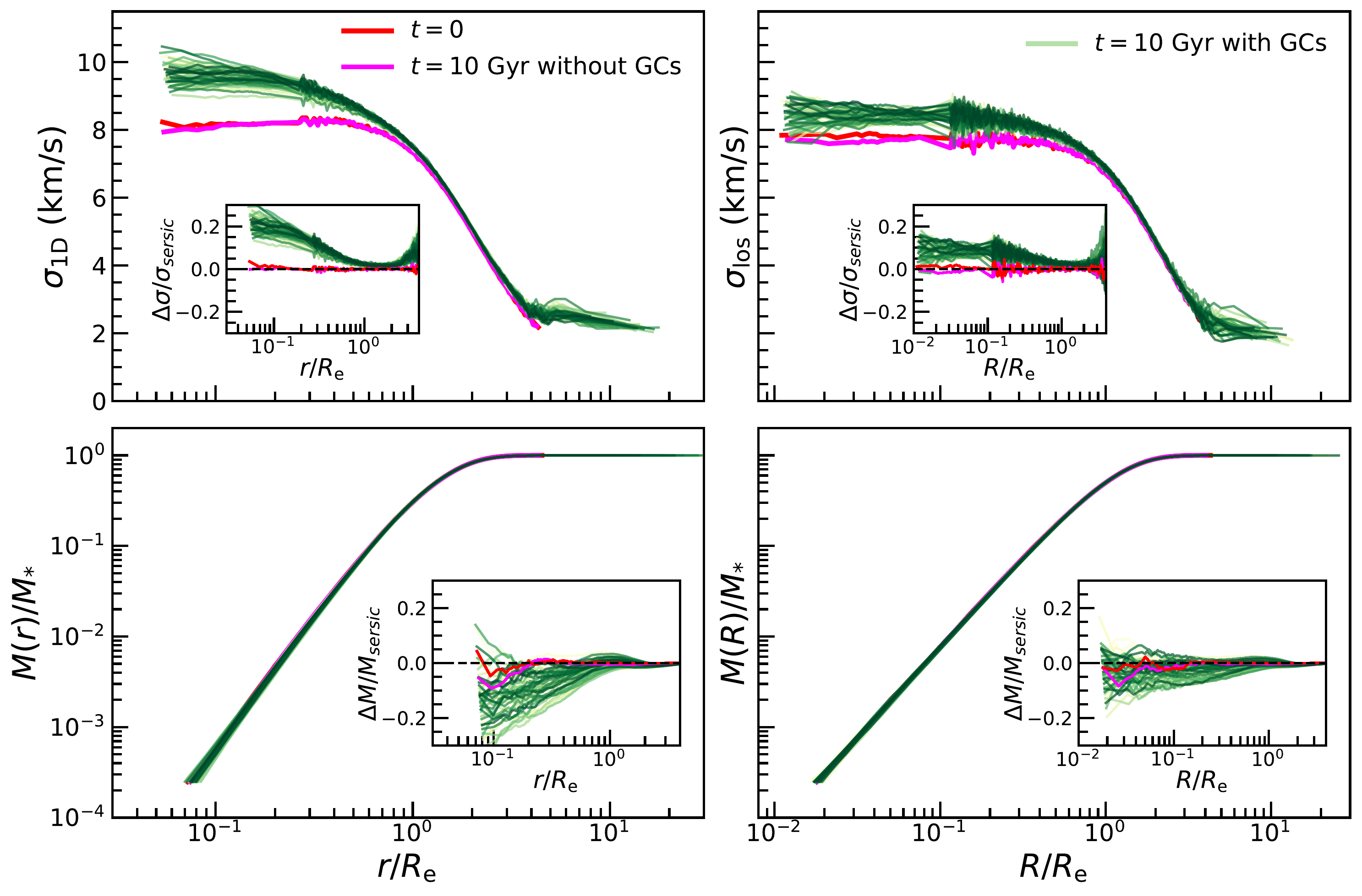}
    \caption{Top-left and bottom-left panels show the 1D velocity dispersion profile ($\sigma_{1D}$) and the 3D enclosed mass profile of the stars at $t=0$ (red curves), at $t=10\ \rm Gyr$ in absence of GCs (magenta curves) and at $t=10\ \rm Gyr$ in presence of GCs (green curves). $\sigma_{1D}$ is calculated by taking the average of the stellar dispersions along x, y, and z axes. Top-right and bottom-right panels show the los velocity dispersion profile and the projected enclosed mass profile of the stars with the same color coding. The magenta curves are in excellent agreement with the red curves, indicating that in the absence of GCs, the stellar body of DF2 is very close to equilibrium. In the presence of multiple GCs, the velocity dispersion is the central region increases and the enclosed mass decreases. The effect is less pronounced in projection. See text for detailed discussion.}
    \label{fig:enclosed_mass_sigma}
\end{figure*}

We now turn to the evolution of the GC system as a whole. The total energy of the GC system defined with respect to the center of mass of all particles (stars+GCs) is given by
\begin{equation}
    E_{\rm tot}=\sum_{i=1}^{N_{GC}} \frac{1}{2} m_{i}v_i^{2} + \sum_{i=1}^{N_{GC}} m_{i} \phi^{*}_{i} - \sum_{i=1}^{N_{GC}} \sum_{\substack{j=1 \\ j<i}}^{N_{GC}} \frac{ G m_{i}m_{j}}{\sqrt{r_{ij}^2+\epsilon^2}}\,.
    \label{E_tot}
\end{equation}
Here, $\vec{v}_{i}$ is the velocity of the $i^{th}$ GC, $m_{i}$ is its mass, $\phi_{i}^{*}$ is the total gravitational potential at its location due to the stellar body, $r_{ij}$ is the distance between the $i^{th}$ and $j^{th}$ GC, $\epsilon$ is the plummer softening used in the simulations and $G$ is the universal gravitational constant. $\phi_{i}^{*}$ is determined by using a tree code with the same opening angle and softening length as used in the simulations. The first term in Equation~\ref{E_tot} is the kinetic energy of the GC system. The second and third terms represent the contributions to the potential energy from the stellar body and from the mutual interactions among the GCs respectively.

Dynamical friction causes $E_{\rm tot}$ to become more negative (i.e., more bound) with time. The drag force causes a reduction in the kinetic energy of the GC system, which causes the GCs to sink towards the center, where the potential energy is more negative. The energy that is released in this process is transferred to the stellar body of DF2, heating it up (see Section~\ref{sec:response}). Figure~\ref{fig:gc_system_energy_change} shows the evolution in $E_{\rm tot}/E_{\rm tot,0}$ for each of our 50 multi-GC simulations. Here $E_{\rm tot,0}$ is the total initial (at $t=0$) binding energy of the GC system. On average, over a period of 10 Gyr, the binding energy of the GC system increases (i.e., becomes more negative) by about 80 percent but with a large realization-to-realization variance.

In the left-hand panels of Figure \ref{fig:gc_cumulative_number_profile}, we show the variance in the 3D cumulative GC number profile ($5^{\rm th}$-$95^{\rm th}$ percentiles) at $t=0$ in blue. The variance in the evolved profile after 2, 5, and 10 Gyr is shown in green. In the right-hand panels of the same figure, the blue histograms depict the cumulative GC number profile in projection at $t=0$, which is same for every realization by construction. The variance in it after 2, 5, and 10 Gyr of evolution is shown in green. During the first 2 Gyr, both the 3D and the projected profiles evolve very little; the change in the projected profile is consistent with no evolution within the realization-to-realization variance. The profiles evolve more over 5 Gyr and are significantly different after 10 Gyr.

In the left and right-hand panels of Figure \ref{fig:gc_rhalf_los_dispersion}, we show the evolution in the projected half-number radius ($R_{\rm half}^{\rm GC}$) and the los dispersion ($\sigma_{\rm los}^{\rm GC}$) of the GC system over 10 Gyr respectively. $R^{\rm GC}_{\rm half}$ is defined here as the mean of the projected radius of the $5^{\rm th}$ and the $6^{\rm th}$ innermost GC and $\sigma_{\rm los}^{\rm GC}$ is the rms spread in the los velocities of the GCs. The medians obtained from the 50 simulations are shown in black at separations of 1 Gyr. The realization-to-realization variance is highlighted by indicating the $16^{\rm th}$-$84^{\rm th}$ percentiles in deep red and $5^{\rm th}$-$95^{\rm th}$ percentiles in light red respectively. The blue and green curves depict the evolution in $R_{\rm half}^{\rm GC}$ and $\sigma_{\rm los}^{\rm GC}$ for two specific realizations: those which have the largest and the smallest $R_{\rm half}^{\rm GC}$ after 10 Gyr of evolution respectively. The GC system in each simulation starts out with the same $R_{\rm half}^{\rm GC}$ and $\sigma_{\rm los}^{\rm GC}$. Dynamical friction causes the GCs to move in and slow down. $R_{\rm half}^{\rm GC}$ evolves more than $\sigma_{\rm los}^{\rm GC}$. The median $R_{\rm half}^{\rm GC}$ and $\sigma_{\rm los}^{\rm GC}$ decrease by $65 \%$ and $34 \%$ respectively over 10 Gyr. The decrease is even more at the $5^{\rm th}$ percentile. However, at the $95^{\rm th}$ percentile, the changes in $R^{\rm GC}_{\rm half}$ and $\sigma_{\rm los}^{\rm GC}$ are consistent with no evolution up to 4-5 Gyr and up to 7-8 Gyr respectively.

From Figures \ref{fig:gc_system_energy_change}, \ref{fig:gc_cumulative_number_profile}, and \ref{fig:gc_rhalf_los_dispersion}, we conclude that while the GC system is definitely affected by dynamical friction over $10 \Gyr$, causing the total energy, cumulative number profile, $R_{\rm half}^{\rm GC}$ and $\sigma_{\rm los}^{\rm GC}$ (at least the median) to change significantly, the changes do not become statistically significant until after about 2 Gyr. This implies that the GCs do not rapidly sink to the stellar core and thus that their phase space coordinates are not unlikely for a baryon-only mass model. However, since the GC system undergoes significant evolution over $10\ \rm Gyr$, it is also clear that if DF2 indeed has no dark matter and formed $\sim 10\ \rm Gyr$ ago, some of the GCs must have formed further out, and the GC system must have been somewhat more extended in the past than what it is today. As a very crude estimate, we can linearly extrapolate the evolution of the median $R_{\rm half}^{\rm GC}$ backwards in time. This suggests that 10 Gyr ago, the median $R_{\rm half}^{\rm GC}$ would have been $5.1 \kpc$, about 2.3 times larger than the stellar half-light radius and about 59 percent larger than its current value of $3.2 \kpc$. Since dynamical friction is less efficient for less bound orbits (i.e., at earlier times), this is likely to be a conservative upper limit on the amount by which $R_{\rm half}^{\rm GC}$ can have evolved over the past $10\ \rm Gyr$.

\begin{figure*}
    \centering
    \includegraphics[width=\textwidth]{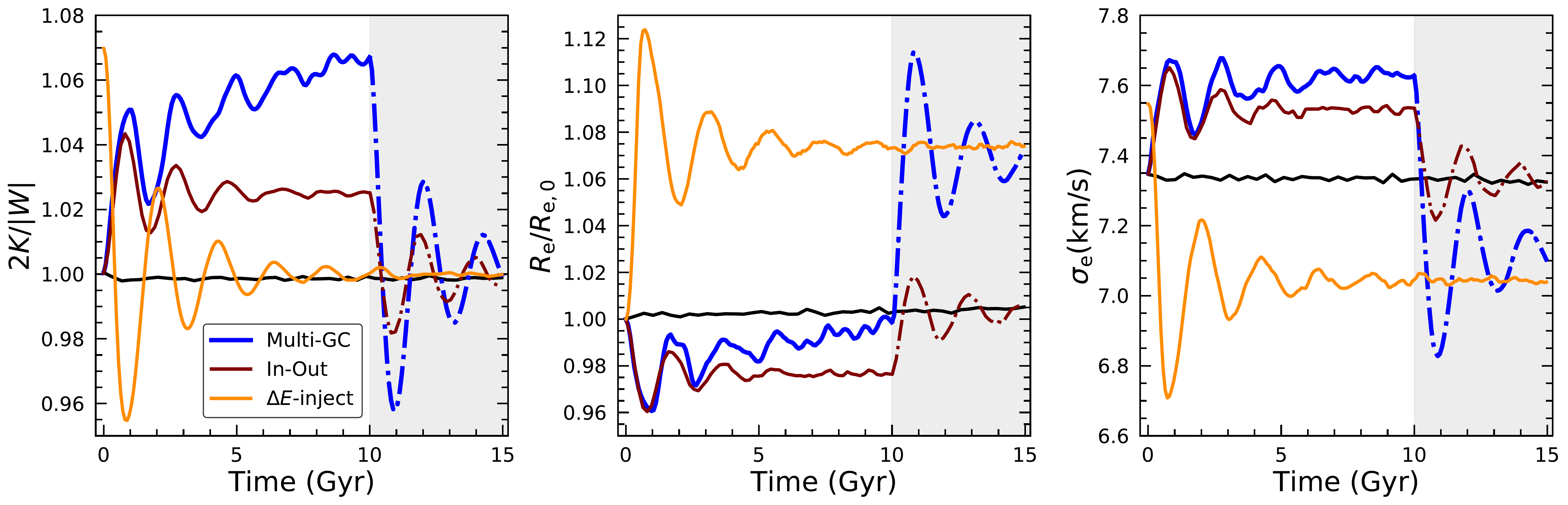}
    \caption{The evolution of the virial ratio of the stars, $2K/|W|$ (left-hand panel), the projected effective radius, $R_\rme$, normalized by its initial value, $R_{\rm e,0}$ (middle panel), and the stellar los velocity dispersion, $\sigma_\rme$ (in $\kms$) within $R_\rme$ (right-hand panel). The blue lines show the mean results obtained from a random subset of 5 of our 50 multi-GC simulations in which we instantaneously remove the GCs at $t=10 \Gyr$ and allow the stellar system to evolve for another 5 Gyr. For comparison, the black lines show the results from our stars-only simulation, which remains in virial equilibrium throughout. The maroon and orange lines show the results from two idealized simulations. The In-Out simulation (maroon lines) shows the impact of instantaneously introducing (at $t=0$) and removing (at $t=10\Gyr$) the GC system, while the $\Delta E$-inject simulation (orange lines) shows how the stellar system reacts to an instantaneous (at $t=0$), homogeneous injection of (kinetic) energy equal to the average amount of energy transferred from the GCs to the stars over a period of 10 Gyr. See text for detailed discussion.}
\label{fig:stellar_heating}
\end{figure*}

\subsubsection{Response of the Stellar System}
\label{sec:response}

We end this section by discussing how the stellar system responds to the evolution of the GCs. As the globular clusters sink towards the core region of DF2, their orbital energy and angular momentum is transferred to the stellar body. The left-hand panels of Figure~\ref{fig:enclosed_mass_sigma} show the 1D velocity dispersion profile (upper panel) and the enclosed mass profile (lower panel) of the stellar body of DF2 in our simulations. The right-hand panels show the same but in projection. The red curves show the initial profiles at $t=0$, while the green curves show the profiles after 10 Gyr of evolution in each of our 50 multi-GC simulations. For comparison, the magenta curves show the results after 10 Gyr of evolution in the stars-only simulation (i.e., without any GCs). Note that the magenta curves are in good agreement with the red curves, indicating that in the absence of GCs, the stellar body of DF2 is very close to equilibrium (see also Appendix~\ref{force_soft}). In the presence of multiple GCs, though, the velocity dispersion within $0.1\ R_{\rme}$ is enhanced by about 20 percent, while the enclosed mass has, in most cases, slightly decreased. Note that the effect is less pronounced in projection, which is to be expected given that the projection operator combines signal from all radii $r \geq R$.

It is tempting to interpret the heating in the central region of DF2 as arising from dynamical friction, which, after all, transfers energy from the GCs to the stars (see Figure~\ref{fig:gc_system_energy_change}). However, since a gravitational system has negative specific heat, injecting heat should cause it to cool down. In particular, if $\Delta E$ is the energy transferred from the GCs to the stars, then we expect the kinetic energy of the stars to reduce, rather than increase by $\Delta E$. The reason is that re-virialization transfers twice the amount of energy added in kinetic form to potential form, causing the system to simultaneously cool and expand \citep[see][]{binney08}. The fact that the central region of DF2 is hotter than before, therefore, seems to imply that the stellar body has not had time yet to re-virialize. However, this is at odds with the notion that dynamical friction is a secular process, which transfers energy from the GCs to the stars slowly over the entire 10 Gyr duration of the simulations. Perhaps, though, the central region is kept out of virial equilibrium due to the `dynamical stirring' coming from the time-variable, combined gravitational potential of the GCs. Although this seems a plausible explanation, there is another aspect of our simulations that may also contribute; as described in Section~\ref{sec:setup}, at $t=0$, the GCs are instantaneously introduced to the stellar system, whose phase-space coordinates were initialized in the absence of any globulars. Hence, the total system (stars + GCs) is initially out of virial equilibrium, and it may well be that the increased velocity dispersion in the central region is merely an outcome of DF2 equilibrating to the presence of the GC system, which, after all, makes up about 4 percent of the total mass. 

In order to test these ideas, we perform a number of additional simulations. For a random subset of 5 of our 50 multi-GC simulations, we instantaneously remove the GCs at $t=10\Gyr$ and evolve the system for an additional 5 Gyr. Figure~\ref{fig:stellar_heating} plots the evolution of the virial ratio, $2K/|W|$, of the stellar body (left-hand panel), its projected effective radius, $R_\rme$, (middle panel) and the projected los velocity dispersion of the stars within that radius, $\sigma_\rme$ (right-hand panel). The period from 10 to 15 Gyr is shaded in grey, highlighting the duration for which the GCs have been removed from the simulations. The blue curves show the average results from our 5 simulations\footnote{All 5 simulations behave very similarly. Therefore, in order to avoid cluttering the figure, we show the mean rather than the individual simulation results}. For comparison, the black lines show the results for our stars-only simulation, which reveal no significant evolution over the full intervel of 15 Gyr. However, as soon as the multi-GC simulations start, the virial ratio of the stars rapidly increases in response to the sudden presence of the GCs. As a consequence of the ensuing re-virialization, $R_\rme$ and $\sigma_\rme$ start to oscillate with a characteristic time scale of $\sim 2 \Gyr$. In addition to oscillating, the virial ratio of the stars also increases slowly with time. After $t=10 \Gyr$, when the GCs are (instantaneously) removed, the virial ratio plummets, and the stellar system undergoes a new set of re-virialization oscillations that clearly bring the system back to virial equilibrium ($2K/|W|=1$). At the same time, $\sigma_\rme$ rapidly decreases, and $R_\rme$ rapidly increases, which reflects the conversion of kinetic to potential energy. 

At the end of the process, the stellar body of DF2 is more extended and colder than it was initially. This is in agreement with the expectations laid out above. However, it remains to be determined whether this is mainly an outcome of the energy transferred to the stars from the GCs as a consequence of dynamical friction or whether it merely reflects a response of the stellar system to the instantaneous introduction (at $t=0$) and instantaneous removal (at $t=10 \Gyr$) of the GCs. In light of the latter, it is important to point out that impulsively cycling matter in and out of the center of a galaxy has a tendency to puff it up; in particular, this mechanism, if repeated multiple times, can create large cores in halos that are initially cusped \citep[e.g.,][]{pontzen12, dutton16}.

In order to shed some light on this conundrum, we run two `idealized' simulations. In the first, we mimic the process of impulsively cycling matter in and out of DF2 as follows. We rerun the stars-only simulation, except that this time, at $t=0$, we instantaneously introduce an analytical potential corresponding to the smooth, average distribution of the GCs, represented by the black, solid curve in the right-hand panel of Figure~\ref{fig:initial_sd_cgn}. The potential is normalized such that the equivalent total mass is equal to the sum of the masses of the 10 GCs (i.e., about 4\% of the total stellar mass of DF2). We then evolve the system for $10 \Gyr$, at which point we instantaneously remove this external potential and continue the simulation for an additional 5 Gyr. The results of this `In-Out' simulation are shown by the maroon curves in Figure~\ref{fig:stellar_heating}. Immediately following the instantaneous introduction of the analytical potential representing the GCs, $2K/|W|$ rapidly increases. After a few oscillations, the virial ratio settles at a value of $\sim 1.025$, reflecting the new, virialized state, which is characterized by a slightly smaller effective radius and a mildly elevated velocity dispersion. The virial ratio deviates from unity since there is an external force on the stellar system. At $t=10\ \rm Gyr$, the analytical potential is instantaneously removed and the stellar system once again re-virializes. Interestingly, it re-virializes to a state that is indistinguishable from the initial state, indicating that the instantaneous introduction and removal of the GCs has no appreciable {\it net} effect on the stellar system. 

For $t \lta 2 \Gyr$, the evolution of $R_\rme$, $\sigma_\rme$ and $2K/|W|$ in the In-Out simulation is very similar to that of our multi-GC simulations, indicating that the early behaviour seen in the latter is governed by the instantaneous introduction of the GCs at $t=0$. At later stages, though, $R_\rme$, $\sigma_\rme$ and $2K/|W|$ in the multi-GC simulations continue to increase at a slow but steady rate, unlike what is seen in the In-Out simulation. This slow increase is due to dynamical friction, which is transferring energy from the GC population to the stars and due to dynamical stirring by the GCs, which keeps the stellar body slightly out of virial equilibrium. Note that unlike the In-Out simulation, the multi-GC simulations do {\it not} return to their initial state after the removal of the GCs. This reflects the impact that dynamical friction has had on the stellar system over the 10 Gyr duration of the simulations. 

The second idealized simulation, in which we instantaneously and uniformly, at $t=0$, inject the energy lost by the GCs due to dynamical friction over a period of 10 Gyr to the stars, further tests this hypothesis. In particular, using Figure~\ref{fig:gc_system_energy_change}, we identify the simulation with the median value of $E_{\rm tot}/E_{{\rm tot},0}$ at $t=10 \Gyr$ and compute the total amount of energy, $\Delta E = |E_{\rm tot} - E_{{\rm tot},0}|$, that the GCs have transferred to the stars over the duration of that simulation. Next, we rerun the stars-only simulation for 15 Gyr, but this time, at $t=0$, we instantaneously increase the speed of each stellar particle such that its total kinetic energy increases by $\Delta E/N_\rmp$. Here $N_\rmp = 10^6$ is the number of stellar particles in the simulation. Hence, this `$\Delta E$-inject' simulation shows how the stellar system reacts to a uniform, impulsive injection of energy $\Delta E$. The results are shown in Figure~\ref{fig:stellar_heating} as the orange curves. Note how the stellar body re-virializes, in about 8 Gyr, to a final state that is remarkably similar to that of our multi-GC simulations. Although dynamical friction is a secular rather than an impulsive process and although dynamical friction will not transfer energy to the stars homogeneously, this supports our conclusion that the difference between the final ($t=15 \Gyr$) and initial ($t=0$) states of the multi-GC simulations reflects the impact of dynamical friction rather than the impact of instantaneously injecting and removing the GCs.

To summarize, the enhancement of the central stellar velocity dispersion in our multi-GC simulations is primarily a consequence of the fact that we have instantaneously introduced the GCs to the stellar system at $t=0$. However, it is also clear that dynamical friction transfers an appreciable amount of energy from the GCs to the stars, which, together with an ongoing dynamical stirring near the core radius, keeps the stellar system out of virial equilibrium. Using the mass estimator of \citet{wolf10}, we ascertain that this can cause an overestimate in the inferred dynamical mass (which typically relies on the assumption of virial equilibrium) by at most 10 percent, and that is after 10 Gyr from today. At present, with the GCs not yet segregated towards the core, their impact is significantly weaker. Hence, we conclude that overall (the evolution of) the GC system only has a very mild impact on the stellar component of DF2.

\begin{figure*}
    \centering
    \includegraphics[width=0.9\textwidth]{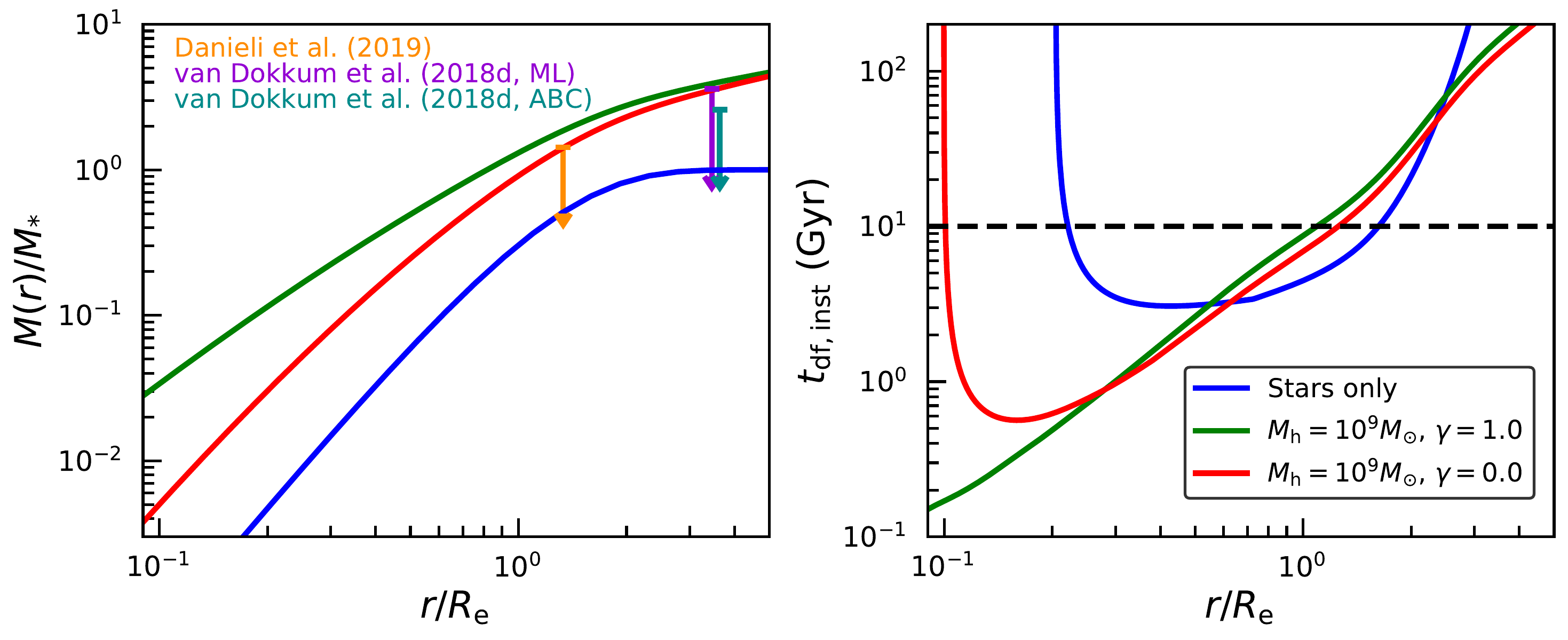}
    \caption{Left-hand panel: The blue curve shows the enclosed mass profile in 3D for the spherically symmetric, baryon-only mass model assumed for DF2 in this study. The total (stars + dark matter) enclosed mass profile in the presence of a standard NFW halo of mass, $M_{h}=10^{9} M_{\odot}$ and inner density slope, $\gamma=1.0$ is shown in green. The same for a cored halo ($\gamma=0$) and identical mass is shown in red. The concentration of the halo in both cases is set according to the average mass-concentration relation of \citet{dutton14}. The three downward pointing arrows highlight the observational constraints on the dynamical mass of DF2 obtained from stellar and GC kinematics \citep{danieli19, vandokkum18c}. Right-hand panel: In the case of a GC of mass, $M_{\rm GC}= 10^{6} M_{\odot}$ moving on a circular orbit, the instantaneous dynamical friction timescale, $t_{\rm df, inst}$ as a function of $r$, the distance from the galaxy center, is shown for each of the three mass models depicted in the left-hand panel. $t_{\rm df, inst}$ is calculated using Equation~\ref{eqn_chandra_df} with radially varying prescriptions for $b_{\rm min}$ and $b_{\rm max}$, as suggested by \citet{just11} and \citet{petts15} respectively. See text for detailed discussion.}
    \label{fig:df_inst_timescale}
\end{figure*}

\section{Summary and Discussion}
\label{sec:summary}

The ultra diffuse galaxy NGC1052-DF2 has an extraordinary population of globular clusters. They are both larger and more luminous than average GCs in the Milky Way, and their total mass is roughly 4 percent of the mass in stars, making DF2 one of the galaxies with the largest specific frequencies known. Most intriguingly, the line-of-sight velocity dispersion of the GCs suggests a total dynamical mass for DF2 that leaves little room for any dark matter \citep[][]{vandokkum18a,vandokkum18c}, a result that is also supported by recently obtained kinematics of the stellar body \citep[][]{danieli19}. Although the notion that this UDG may be entirely devoid of dark matter is contentious, current kinematic data cannot rule it out.

In this paper, we have investigated the feasibility of such a baryon-only model by examining whether in the absence of dark matter, the observed GC population experiences rapid orbital decay due to dynamical friction. Using the Sersic profile that best fits the observed surface brightness distribution, we construct a spherically symmetric mass model for DF2 and run a suite of 50 multi-GC $N$-body simulations. In each of these simulations, the initial positions and velocities of the stars are sampled from an ergodic distribution function corresponding to the assumed mass model. The initial phase-space coordinates of the GCs match constraints on their projected positions and los velocities. Their positions along the los and velocity components perpendicular to the los are sampled from a spherically symmetric, isotropic distribution function constructed by assuming the GC system to be in equilibrium with the baryonic potential. 

We find that the uncertainty in the phase-space coordinates of the GCs translates to a substantial amount of realization-to-realization variance in their evolution. On average, though, the five innermost GCs in projection at $t=0$ experience significant orbital evolution due to dynamical friction; they spiral towards the core radius of DF2 (roughly at $0.2-0.3\ R_\rme$), where they experience core-stalling. As multiple GCs start to congregate near that radius, they experience significant GC-GC interactions, which gives rise to an additional, effective buoyancy preventing them from sinking to the center of the galaxy. In addition to the GC-GC interactions, the presence of multiple GCs also implies that the stellar body of DF2 is constantly being `stirred' by the multiple response densities associated with each of the GCs. This stirring acts like a form of violent relaxation (a rapid time-variability of the gravitational potential), which also contributes to the dynamical buoyancy of the globular cluster population, i.e., violent relaxation acts to negate the mass-segregation arising from dynamical friction. As a consequence, we find that none of the GCs ever manage to sink all the way to the center. 

The five outermost GCs, on average, experience significantly less dynamical friction, and typically only lose a small fraction of their orbital angular momentum within a Hubble time. They rarely interact with each other, or with the other GCs, and they continue to orbit at relatively large distances from the centre. Over a period of 10 Gyr, the median half-number radius of the GC system shrinks by $65\%$ and the median los velocity dispersion of the GCs decreases by $34\%$. In the same time interval, the median total energy of the GC system becomes more negative by a factor of $1.8$. Thus, the GC system becomes more bound, compact and slightly colder with time, but this becomes statistically significant only after about 2 Gyr. We, therefore, conclude that while the current phase-space coordinates of the GCs are not inconsistent with a baryon-only model, the GC system as a whole must have been somewhat more extended in the past compared to what is observed today.

This conclusion is somewhat at odds with a recent study by \citet{nusser18a} who, also based on $N$ body simulations, argues that a dark-to-stellar mass ratio of at least a few tens is required to explain the presence of old, massive GCs in DF2. For a smaller halo mass, \citet{nusser18a} finds that a typical GC spirals into the center of the galaxy in less than 10 Gyr, a timescale that is smaller than the average age of the GCs.  However, \citet{nusser18a} only examined orbits with one particular value of initial eccentricity, considered GC masses that are (somewhat) too large and did not account for GC-GC interactions. In addition, the simulations of \citet{nusser18a} reveal a rather peculiar orbital decay, whereby only the apocenter decreases with time, while the pericenter remains almost fixed; i.e., the orbit circularizes, similar to what happens in the case of grazing encounters \citep[see e.g.,][]{bontekoe87, vandenbosch99}. This is a consequence of the unrealistic initial conditions adopted by \citet{nusser18a}, where all particles (other than the GC) start out with a speed equal to the local circular speed. As a result, at pericenter, the globular cluster is moving faster than all local field particles, and $\rho(<v_S)$ is close to the total density of field particles. At apocenter, though, the globular cluster is moving slower than all local particles, and $\rho(<v_S) \simeq 0$. Hence, dynamical friction is strongly suppressed near apocenter and close to maximally efficient near pericenter (see Equation ~\ref{eqn_chandra_df}), giving rise to a dramatic circularization of the GC orbit. In our simulations, the initial velocities of the stars are properly sampled from an ergodic distribution function corresponding to the assumed mass model, and our simulations do not reveal a similar orbital circularization. 

More importantly, \citet{nusser18a} exclusively considered dark matter halos with a steep central cusp. Consequently, there is no core-stalling in his simulations. Assuming a baryon-only model for DF2, we find that core-stalling and GC-GC interactions give rise to a dynamical buoyancy that prevents the GCs from reaching the very centre of the galaxy, where they would otherwise merge to form a nuclear star cluster. Hence,  contrary to \citet{nusser18a}, we conclude that a high halo mass is {\it not} a necessary requirement to explain the presence of the GC population in DF2 as long as we allow for the GC system to have been somewhat more extended in the past.

In our analysis, we have ignored the tidal field of the nearby massive elliptical galaxy, NGC 1052. DF2 shows no obvious evidence of tidal disturbances, with regular isophotes out to $2\ R_{\rme}$ ($4.4 \ \rm kpc$); constraining the tidal radius to be $ \gtrsim 4\ \rm kpc$ \citep[see also][]{washerman18}. Therefore, in its current configuration and future evolution, as studied here, the GC system is unlikely to be affected by tides. However, if some of the GCs formed further out, then it is possible (although not certain) for them to have been tidally perturbed. While tidal heating will increase the orbital energy of the GCs and negate the effect of dynamical friction, too much energy transfer can also unbind them from the galaxy. In the latter scenario, to account for its currently observed specific frequency in bound GCs, DF2 would have had to start out with an even richer GC population. A second caveat of our analysis is related to the modeling of the GCs as rigid Plummer spheres. In doing so, we have ignored the possibility of GC-GC mergers, which may be relevant as the internal velocity dispersion of GCs is of the same order as the velocity dispersion of the GC system. We intend to investigate the potential impact of GC-GC mergers in a future study.

While this paper has explored the evolution of the GC system for a baryon-only model, current constraints on the dynamical mass of DF2 also allow for the presence of dark matter. This is evident from the left-hand panel of Figure \ref{fig:df_inst_timescale}, which plots the stellar mass enclosed within a 3D radius, $r$, for the spherically symmetric, baryon-only model of DF2 adopted in this study (blue curve). The green and red curves show the total (stars + dark matter) enclosed mass profiles in the presence of a dark matter halo of mass $M_\rmh = 10^9 \Msun$; in the case of the green curve we adopt a standard NFW density profile with an inner density slope of $\gamma \equiv \rmd\log\rho/\rmd\log r = -1$, while the red curve has $\gamma = 0$. In both cases, the concentration of the halo is taken from the average mass-concentration relation of \citet{dutton14}, which yields a scale radius of $\sim 1.3 \kpc$ ($\sim 0.6\ R_\rme$). 

For comparison, the three downward pointing arrows indicate observational constraints on the dynamical mass of DF2. The orange arrow indicates the upper limit (at 95\% confidence) obtained from the stellar velocity dispersion measured by \citet{danieli19} using the Wolf estimator \citep{wolf10}, while the dark cyan and magenta arrows mark the upper limits (at 90\% confidence) inferred from the los velocity dispersion of the GCs measured by \citet{vandokkum18c} using Approximate Bayesian Computation (ABC) and Gaussian maximum likelihood (ML) combined with the Tracer Mass Estimator of \citet{watkins10} respectively. As is evident, all three profiles are more or less consistent with these existing kinematic constraints.
  
We can get some insight as to how the presence of a dark matter halo impacts our results by considering the radial dependence of the {\it instantaneous} dynamical friction time scale, $t_{\rm df,inst}(r) \equiv v_\rmc(r)/a_{\rm df}(r)$. Here $v_\rmc(r) = \sqrt{G M(r)/r}$ is the circular velocity at radius $r$, and $a_{\rm df}(r)$ is the local deceleration due to dynamical friction. We compute  $t_{\rm df,inst}(r)$ for each of the three mass models shown in the left- hand panel of Figure \ref{fig:df_inst_timescale} adopting a GC mass of $M_{\rm GC}= 10^{6} M_{\odot}$ and calculating $a_{\rm df}(r)$ using Chandrasekhar's dynamical friction formula (Equation~\ref{eqn_chandra_df}). The density of background particles moving slower than $v_\rmc$, $\rho(<v_\rmc$), is evaluated using ergodic distribution functions corresponding to the assumed stellar plus dark matter density profiles. In addition, we follow \citet{just11} and \citet{petts15} and compute the Coulomb logarithm, $\ln\Lambda = \ln(b_{\rm max}/b_{\rm min})$, assuming radially dependent maximum and minimum impact parameters. In particular, we use $b_{\rm max}(r) = {\rm min}\left(r, r/|\rmd\log\rho/\rmd\log r|\right)$, and $b_{\rm min}(r) = G M_{\rm GC}/v^2_\rmc(r)$. As the subject mass sinks towards the center, $b_{\rm max}$ decreases. Once it becomes comparable to $b_{\rm min}$, then $\rm{ln} \Lambda \to 0$ and dynamical friction ceases. For a circular orbit, this happens when the enclosed mass $M(r) = M_{\rm GC}$. We note that this prescription slightly underestimates the stalling radius as it does not take into account the effect of resonances (see Section~\ref{sec:corestalling}).

The results are shown in the right-hand panel of Figure \ref{fig:df_inst_timescale}. In the absence of dark matter, $t_{\rm df,inst}$ decreases with decreasing radius, reaching a nearly constant minimum of $\sim 3 \Gyr$ between roughly $0.3\ R_\rme$ and $0.6\ R_\rme$. For $r \lta 0.3\ R_\rme$, it rapidly increases with decreasing radius, with $t_{\rm df,inst} \to \infty$ as $r$ approaches $\sim 0.2\ R_\rme$. This is roughly the radius where the enclosed stellar mass is equal to the mass of the GC and core-stalling takes effect. At $r \gta 1.5\ R_\rme$, the instantaneous dynamical friction time exceeds the Hubble time, and dynamical friction is inefficient as well. Hence, one expects that GCs that start out at $r \gta 1.5\ R_\rme$ experience little to no dynamical friction over a Hubble time, while those with an initial radius $r \lta 1.5\ R_\rme$ sink towards $r \sim 0.3\ R_\rme$ within about 3 Gyr. This is in qualitative agreement with the behaviour seen in our simulations.

In the presence of a NFW dark matter halo of mass $10^{9} M_{\odot}$, which is allowed by current kinematic constraints, $t_{\rm df,inst}$ is slightly larger than in the absence of dark matter at intermediate radii ($0.5 \lta r/R_\rme \lta 2$), indicating that dynamical friction is less effective over that radial range. However, at $r \lta 0.5\ R_\rme$, the instantaneous dynamical friction time scale is substantially smaller than that in the case of a baryon-only mass model\footnote{This is also the case for $r \gta 2.5\ R_\rme$; however, this is of no significant consequence as the time scales are too long for any appreciable amount of orbital decay within a Hubble time.}. In fact, since $t_{\rm df,inst}$ rapidly declines with decreasing radius, we conclude that in the presence of a steep $r^{-1}$ density cusp, characteristic of a NFW halo, all GCs that start out at an initial radius $r \lta R_\rme$ will spiral all the way to the centre of DF2 within a Hubble time, where they are likely to merge and form a nuclear star cluster. If the dark matter halo instead has a central core, $t_{\rm df,inst}$ reaches a minimum at a radius where the total enclosed mass is comparable to that of the globular, where we thus expect the globulars to pile up (in the absence of GC-GC interactions). In the case depicted in Figure~\ref{fig:df_inst_timescale} (red line), this stalling radius is roughly $0.1\ R_\rme$, about a factor two smaller than the stalling radius in the absence of dark matter. 

Based on these calculations, we speculate that as long as a potential dark matter halo has a sufficiently large, roughly constant density core, core-stalling and GC-GC interactions are likely to prevent the GCs from sinking too rapidly to the center. If, however, the halo has a NFW-like cusp, we expect that the globulars will sink to the center where they are likely to form a nuclear star cluster, unless the mass of the halo is sufficiently large that the time scale for this to happen is too long. The results by \citet{nusser18a} suggest that this requires a dark-to-stellar mass ratio of at least a few tens. Since such a massive halo is difficult to reconcile with existing kinematic constraints (see left-hand panel of Figure~\ref{fig:df_inst_timescale}), we conclude that if DF2 has a dark matter halo, it requires a significant core. In this respect, DF2 resembles the well-studied Fornax dwarf galaxy. Similar to DF2, Fornax has a relatively large population of old GCs for its stellar mass. Standard arguments suggest that these GCs should sink to the center via dynamical friction in much less than a Hubble time \citep[e.g.,][]{oh00}. Their presence and the absence of a nuclear star cluster has, therefore, been used to argue that Fornax must have a cored dark matter distribution \citep[e.g.,][]{hernandez98, read06, goerdt06, inoue09, cole12, arca-sedda16}. The main difference with DF2, though, is that whereas the kinematics of Fornax clearly requires the presence of dark matter \citep[e.g.,][]{strigari06}, DF2 seems to prefer none or very little. It remains to be seen to what extent baryonic processes \citep[e.g,][]{pontzen12, dutton16} or tidal stripping \citep[][]{ogiya18} can produce systems like DF2 with little dark matter and overly massive populations of globular clusters. For the moment, based on our results, we conclude that a baryon-only model for DF2 is not inconsistent with the data.

\acknowledgments

DDC acknowledges Marla Geha for a stimulating discourse on DF2 in her stellar dynamics class, which largely prompted this study. We are grateful to Uddipan Banik for many insightful discussions, to Nir Mandelker, Seshadri Sridhar, and Scott Tremaine for comments and advice and to the anonymous referee for valuable feedback. FvdB is supported by the National Aeronautics and Space Administration through Grant No. 17-ATP17-0028 issued as part of the Astrophysics Theory Program, and receives additional support from the Klaus Tschira foundation, and from the US National Science Foundation through grant AST 1516962. PvD gratefully acknowledges support from STScI grant HST-GO-14644.

\appendix

\section{Choice of Gravitational Force Softening}\label{force_soft}

The main goal of this paper is to examine how the population of GCs in DF2 evolves under the influence of dynamical friction. Hence, it is prudent that we capture the impact of dynamical friction as accurately as possible, and this is the main principle that guides us in choosing the softening length, $\epsilon$. The dynamical friction force, as envisioned in \cite{chandrasekhar43}, is considered as a sum of uncorrelated two-body encounters between the subject mass (here one of the GCs in DF2) and the individual background particles (here a stellar particle of DF2). Under the assumption of an infinite and homogeneous distribution of background particles, the deceleration due to dynamical friction is proportional to
\begin{equation}
\calI \equiv \int_{b_{\rm min}}^{b_{\rm max}} \Delta v_{\rm rel}(b,v_{\infty}) \, b \, \rmd b\,.
\label{Idf}
\end{equation}
Here $b_{\rm min}$ and $b_{\rm max}$ are the minimum and maximum impact parameters, and $\Delta v_{\rm rel}(b,v_{\infty})$ is the change in the relative velocity parallel to the initial direction of motion experienced by the reduced particle moving in a straight line orbit with an impact parameter, $b$, and a relative velocity, $v_{\infty}$. This change in the relative velocity depends on the force field of the GC and hence on its density profile. In reality, a GC is well represented by a King profile \citep[][]{king62, king66}. In our simulations, though, we represent them using Plummer spheres of characteristic radius equal to the softening length. Let $\calI_{\rm King}$ be the value of $\calI$ obtained from Equation~(\ref{Idf}) for a realistic GC with a King profile of realistic characteristic size, and let $\calI_{\rm Plummer}(\epsilon)$ correspond to the value of $\calI$ for a Plummer sphere of characteristic radius equal to $\epsilon$. We tune $\epsilon$ such that $\calI_{\rm Plummer}(\epsilon) = \calI_{\rm King}$ as follows.

\begin{figure}[t]
    \centering
    \includegraphics[width=0.9\textwidth]{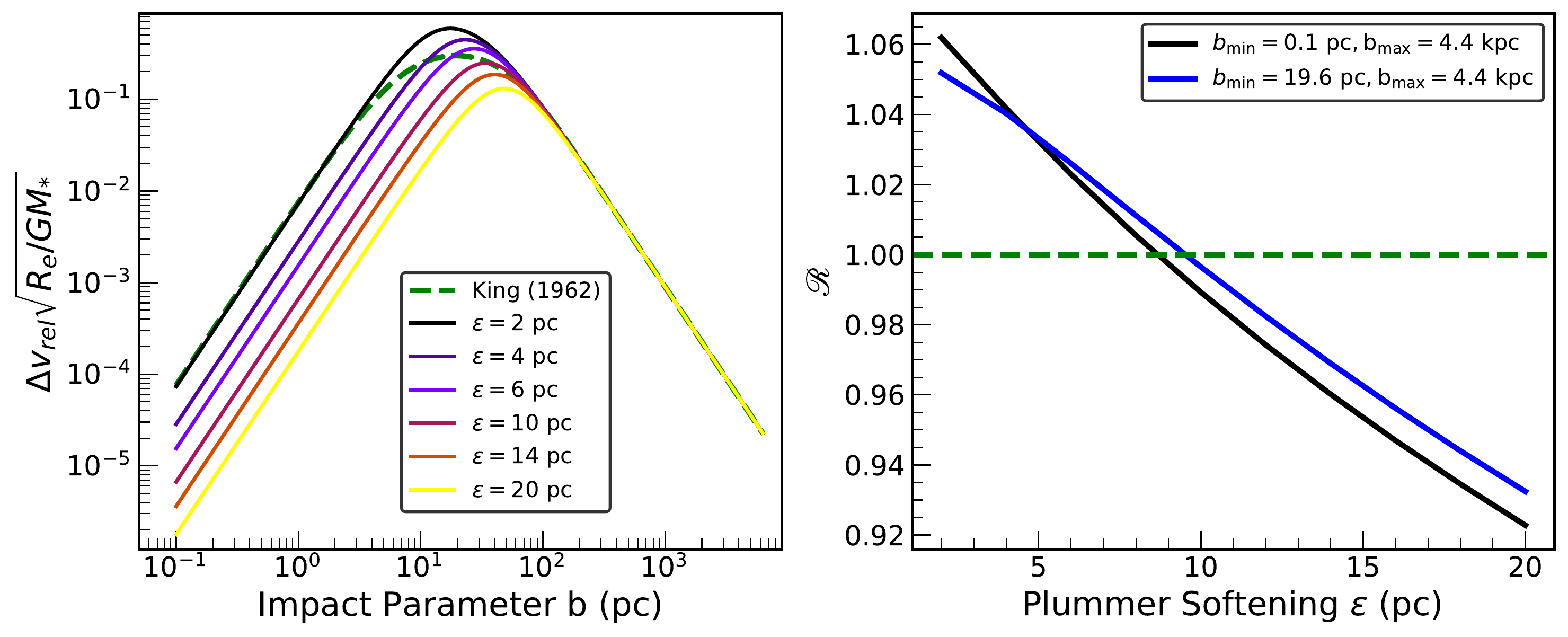}
    \caption{Left-hand panel: Change in the relative velocity parallel to the initial direction of motion, $\Delta v_{\rm rel}$, as a function of the impact parameter, $b$, experienced by the reduced particle moving in a straight line orbit with a relative velocity of $v_{\infty}=9\ \rm km/s$. The green, dashed line is for a \citet{king62} sphere of mass, $\bar{M}=6.7 \times 10^{5}\ \rm M_{\odot}$, tidal radius, $\bar{r}_t=122.7\ \rm pc$ and core radius, $\bar{r}_c=1.2\ \rm pc$. The solid lines are for Plummer spheres of same mass but different characteristic sizes, $\epsilon$. Right-hand panel: $\calR$ as a function of $\epsilon$ where $\calR$ is defined according to Equation \ref{calR}. Black and blue curves are for $b_{\rm min}=0.1\ \rm pc$ and $19\ \rm pc$ respectively. The optimal softening is the $\epsilon$ for which $\calR=1$ and it turns out to be $\sim 10\ \rm pc$.} 
    \label{fig:softening}
\end{figure}

The average GC in DF2 has a $V$-band magnitude of $\bar{M}_V = -9.2$ and a half-light radius of $\bar{r}_\rmh = 6.2 \pc$ \citep[][]{vandokkum18b}. Under the assumption of a $V$-band mass-to-light ratio of $1.8$, it has a mass of $\bar{M}=6.7 \times 10^{5} M_{\odot}$. If we adopt a central surface brightness of $\bar{\Sigma}_{0} = 1.2 \times 10^{4} \Lsun \pc^{-2}$, which is equal to the mean central surface brightness of Milky Way GCs above $M_{V}=-8.6$ \footnote{Based on the \citet{harris96} catalog of Milky Way GCs.} (the luminosity of the faintest DF2 GC), and assume that the GCs follow a \citet{king62} profile, we infer a core radius of $\bar{r}_\rmc = 1.2 \pc$ and a tidal radius of $\bar{r}_\rmt = 122.7 \pc$. We use these parameters and an orbit integrator to compute $\Delta v_{\rm rel}(b)$ for a typical GC-star encounter velocity of $v_\infty \sim \sqrt{2} \, \langle \sigma_{*} \rangle = 9 \kms$, where $\langle \sigma_{*} \rangle = 6.5\kms$ is the total los velocity dispersion of the stars in DF2, obtained from the stellar distribution function (see Section \ref{sec:setup}). The result is the green, dashed line in the left-hand panel of Figure~\ref{fig:softening}.

Next, we use the same orbit integrator to compute $\Delta v_{\rm rel}(b)$ for a Plummer sphere of the same mass but with a characteristic radius equal to $\epsilon$, whose force field is given by $F_{\rm Plummer} = G \bar{M} r/(r^2 + \epsilon^2)^{3/2}$. Throughout, we adopt star particles with a mass of 200 $M_{\odot}$, which is equal to the mass of the star particles in our simulations (see Section~\ref{sec:setup}). The solid lines in the left-hand panel of Figure~\ref{fig:softening} show the results for different values of $\epsilon$. At large impact parameters, $\Delta v_{\rm rel} \propto b^{-2}$, as expected for a point mass \citep[][]{chandrasekhar43}. However, when $b$ becomes comparable to the characteristic radius of the GC, the detailed mass profile of the GC becomes important. In the limit $b \to 0$, the enclosed mass $\bar{M}(<b)$ goes to zero, as does $\Delta v_{\rm rel}$. Hence, the maximum deceleration arises for an impact parameter that is comparable to the characteristic radius of the GC.

Using the results shown in the left-hand panel of Figure~\ref{fig:softening}, we now compute
\begin{equation}
\calR(\epsilon) \equiv \calI_{\rm Plummer}(\epsilon)/ \calI_{\rm King} \,, 
\label{calR}
\end{equation}
which we plot in the right-hand panel of the same figure. The two different curves correspond to different values for the minimum impact parameter used in the integration (see Equation~\ref{Idf}): $b_{\rm min} = 0.1 \pc$ (black curve), which is 10 percent of $\bar{r}_\rmc$, and $b_{\rm min} = 19.6 \pc$ (blue curve), which is the impact parameter for which $\Delta v_{\rm rel}(b)$ of the King profile is maximum. In both cases, we adopt  $b_{\rm max} = 4.4 \kpc$, which is equal to two times the projected effective radius, $R_\rme$ of the stars in DF2. Note that the choice of $b_{\rm min}$ only has a mild impact on $\calR(\epsilon)$. If we now define the optimal softening length according to $\calR(\epsilon_{\rm opt}) = 1$, we obtain  $\epsilon_{\rm opt} = 8.7\pc$ ($9.5 \pc$) for $b_{\rm min} = 0.1 \pc$ ($19.6 \pc$). We emphasize that these results are insensitive to the exact value of $b_{\rm max}$ and to the mass of the star particle as long as the latter is significantly smaller than that of the GC. This implies that the value of $\epsilon_{\rm opt}$ is independent of the number of star particles used in the simulations. Although this derivation of the optimal force softening is based on a number of oversimplified assumptions (straight orbits, uniform distribution of impact parameters, fixed $v_{\infty}$, etc.), it is reassuring that previous investigations of core stalling based on $N$-body simulations \citep[e.g.,][]{read06,cole12} have indeed adopted a force softening of 5-10 pc for particles representing GCs.

The above derivation of the optimal force softening only considers GC-star interactions. However, force softening is also important for the dynamics of the stars in DF2 as a softening length that is too small can result in significant two-body relaxation. In order to test the impact of force softening on the evolution of the stellar body, we run a number of $N$-body simulations with only stars (i.e., without GCs) using softening lengths spanning the entire range from $5\pc$ to $1100 \pc$ (50\% of $R_\rme$). Since DF2 has a central density core, the results are extremely insensitive to the value of $\epsilon$; we find that the stellar body remains in stable equilibrium for more than a Hubble time for $5 \pc < \epsilon < 220 \pc$. Hence, a softening of $\epsilon=10\pc$ is also adequate to resolve the dynamics of the stars in DF2. Indeed, as is evident from the red and magenta curves in Figure \ref{fig:enclosed_mass_sigma}, in the absence of GCs, the density and velocity structure of DF2 shows no significant evolution over the duration of the simulation (10 Gyr).

Based on these considerations, we adopt a force softening of $\epsilon=10\pc$ throughout.

\section{Choice of Time Step}\label{time_step}

\begin{figure}
    \centering
    \includegraphics[width=0.9\textwidth]{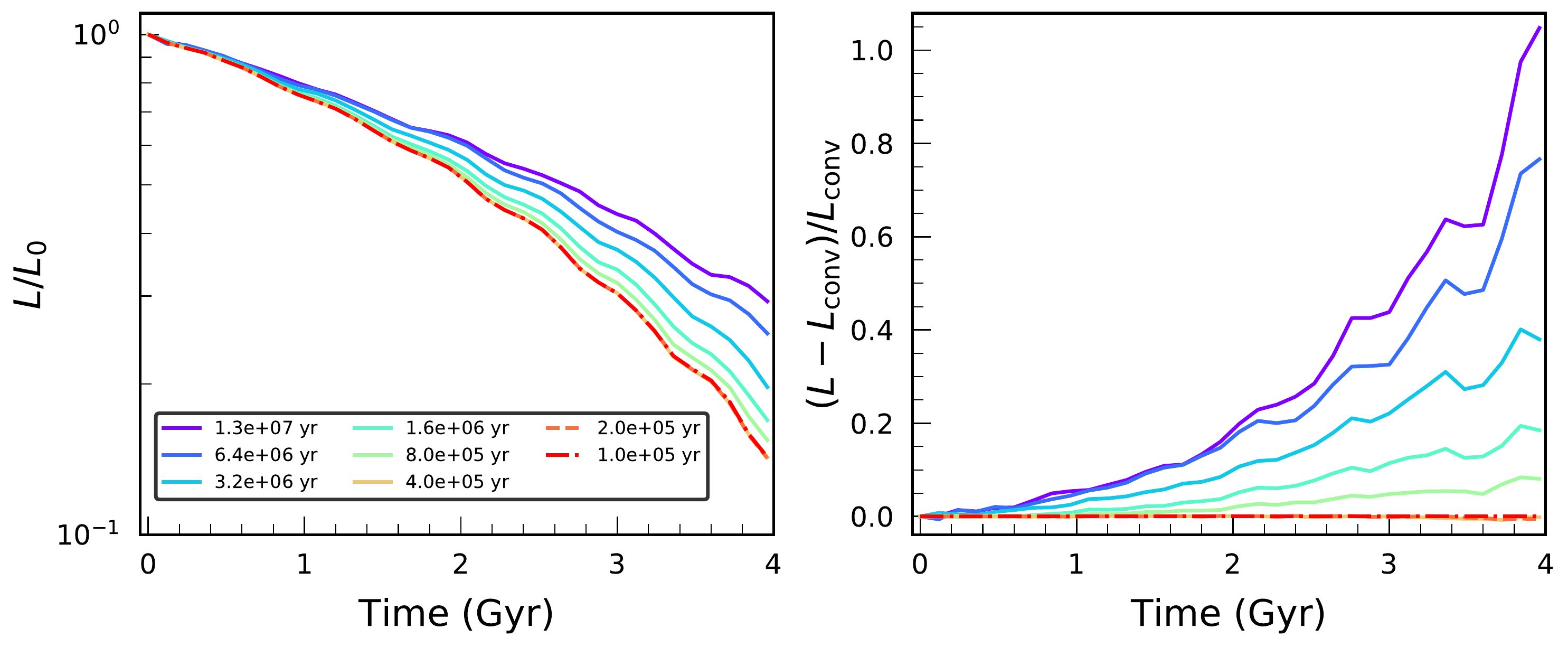}
    \caption{Left-hand panel: Evolution of the GC angular momentum in test runs with $10^{6}$ star particles and the most massive GC. The initial conditions for the GC are taken from one of the 50 realizations. The softening length is 10 pc and the time step for each run is indicated in the legend. Right-hand panel: The percent difference in the angular momentum of the GC in each run with respect to the angular momentum in the run that has a time step of $1 \times 10^{5} \rm yr$ (denoted by $L_{\rm conv}$). Runs with a time step of $\Delta t \le 4 \times 10^{5} \yr$ are converged to better than 1 percent. However, to keep the computational time manageable, we choose a time step of $8 \times 10^{5}\ \rm Gyr$ for all simulations with GCs.}
    \label{fig:time_step}
\end{figure}

The time step for the stars-only simulation is chosen as $1.7 \times 10^{7}\ \rm yr$. This is equal to $(1/50)^{\rm th}$ of the orbital time for a circular orbit at $0.1 R_{\rme}$, which is well inside the core region of the galaxy. For the runs with GCs, the time step is determined by running a series of test simulations having $10^{6}$ star particles and the most massive GC. The initial conditions for the GC are taken from one of the 50 realizations. The first simulation of the series has a time step of $1.3 \times 10^{7}\ \rm yr$. In each subsequent simulation, the time step is reduced by a factor of $\sim 2$. The series is continued until a convergence in the angular momentum evolution of the GC is attained. For all runs, we adopt the same force softening as that in the science runs ($\epsilon=10$pc).

In Figure~\ref{fig:time_step}, we show the evolution of the GC angular momentum for each simulation of the series. Runs with a time step of $\Delta t \le 4 \times 10^{5} \yr$ are converged to better than 1 percent. Unfortunately, keeping the total computational time for our large set of simulations manageable requires $\Delta t \gta 8 \times 10^{5} \yr$. As a compromise, we, therefore, adopt $\Delta t = 8 \times 10^5 \yr$ throughout. As is apparent from Figure \ref{fig:time_step}, at this temporal resolution we slightly overestimate the time it takes for the globular to lose its angular momentum, but the effect is sufficiently small that it does not significantly impact any of our main conclusions.

\bibliography{ms}
\bibstyle{aasjournal}
\end{document}